\documentclass[twocolumn]{aastex631}

\usepackage{amsmath}
\usepackage{CJKutf8}

\received{xxx}
\revised{xxx}
\accepted{xxx}
\published{xxx}

\submitjournal{ApJS}

\begin{document}
\begin{CJK*}{UTF8}{gbsn}

\newcommand{\hw}[1]{\textcolor{red}{[HW: #1]}}
\newcommand{\dg}[1]{\textcolor{blue}{[DG: #1]}}

\title{\texttt{jetsimpy}: A Highly Efficient Hydrodynamic Code for Gamma-Ray Burst Afterglow}

\correspondingauthor{Hao Wang}
\email{wang4145@purdue.edu}

\author[0000-0002-0556-1857]{Hao Wang(王灏)}
\affiliation{Department of Physics and Astronomy, Purdue University, 525 Northwestern Avenue, West Lafayette, IN 47907, USA}

\author{Ranadeep G. Dastidar}
\affiliation{Department of Physics and Astronomy, Purdue University, 525 Northwestern Avenue, West Lafayette, IN 47907, USA}

\author[0000-0003-1503-2446]{Dimitrios Giannios}
\affiliation{Department of Physics and Astronomy, Purdue University, 525 Northwestern Avenue, West Lafayette, IN 47907, USA}

\author{Paul C. Duffell}
\affiliation{Department of Physics and Astronomy, Purdue University, 525 Northwestern Avenue, West Lafayette, IN 47907, USA}

\begin{abstract}
Gamma-ray burst (GRB) afterglows are emissions from ultrarelativistic blast waves produced by a narrow jet interacting with surrounding matter. Since the first multimessenger observation of a neutron star merger, hydrodynamic modeling of GRB afterglows for structured jets with smoothly varying angular energy distributions has gained increased interest. While the evolution of a jet is well described by self-similar solutions in both ultrarelativistic and Newtonian limits, modeling the transitional phase remains challenging. This is due to the nonlinear spreading of a narrow jet to a spherical configuration and the breakdown of self-similar solutions. Analytical models are limited in capturing these nonlinear effects, while relativistic hydrodynamic simulations are computationally expensive, which restricts the exploration of various initial conditions. In this work, we introduce a reduced hydrodynamic model that approximates the blast wave as an infinitely thin two-dimensional surface. Further assuming axial symmetry, this model simplifies the simulation to one dimension and drastically reduces the computational costs. We have compared our modeling to relativistic hydrodynamic simulations and semianalytic methods, and applied it to fit the light curve and flux centroid motion of GRB 170817A. These comparisons demonstrate good agreement and validate our approach. We have developed this method into a numerical tool, \texttt{jetsimpy}, which models the synchrotron GRB afterglow emission from a blast wave with arbitrary angular energy and Lorentz factor distribution. Although the code is built with GRB afterglow in mind, it applies to any relativistic jet. This tool is particularly useful in Markov Chain Monte Carlo studies and is provided to the community.
\end{abstract}


\section{Introduction} \label{sec:intro}
Gamma-ray bursts (GRBs) are the most energetic catastrophic events in the Universe. Their tremendous explosive power provides a platform for the scientific study of fundamental physical processes in extreme physical environments. One of the intriguing aspects of GRBs is their long-term, multiwave band afterglow following the prompt emission. GRB afterglows are radiation from the ultrarelativistic blast waves produced by the interaction of a narrow jet with the surrounding external medium. They are the promising sites to study the hydrodynamical interactions and microphysical processes associated with some of the most extreme macroscopic bulk motions in the Universe \citep{1993ApJ...405..278M,1999ApJ...517L.109S}. Recently, it has also been demonstrated that GRB afterglows could serve as probes to study the Universe's expansion rate (see \citealt{2022Univ....8..289B} for a review), thereby helping to unravel the nature of the Hubble tension (see \citealt{2021CQGra..38o3001D} for a review).

To model the evolution of a narrow GRB jet, the angular energy profile is often approximated by a so-called ``top-hat" model, where the energy is uniformly distributed within a narrow core, and sharply drops to zero at the edge. This model has been proven highly successful in accounting for the GRB phenomenology when the jet is observed from an on-axis direction, namely when the observer's line of sight (LOS) lies within the jet core (see \citealt{2015PhR...561....1K} for a review). This is a natural consequence of the selection bias, because at cosmological distances it is only possible to observe a GRB within the jet core, where the prompt emission is bright enough for detection (e.g., \citealt{2019MNRAS.482.5430B}). In this situation, the less energetic tail outside the core becomes subdominant in observations, making the study of a smoothly varying angular energy profile less compelling.

The revolutionary discovery of
the first multimessenger observation of a binary neutron star merger (\citealt{2017PhRvL.119p1101A, 2017ApJ...848L..12A, 2017ApJ...848L..13A}; see \citealt{2021ARA&A..59..155M} for a review), the famous gravitational wave event GW170817 and GRB 170817A, has revealed the importance of a more sophisticated jet model: the so-called ``structured jet" (e.g. \citealt{1998ApJ...499..301M,2002MNRAS.332..945R,2003ApJ...591.1086G, 2003ApJ...591.1075K}). A structured jet is characterized by the previously mentioned smoothly varying angular energy profile, in which a less energetic tail extends far beyond the core. This model is motivated by the argument that the afterglow of GRB 170817A was very likely observed from an off-axis direction \citep{2017ApJ...848L..25H, 2017ApJ...848L..20M, 2018ApJ...856L..18M, 2017Natur.551...71T, 2018MNRAS.478L..18T, 2020MNRAS.498.5643T, 2018ApJ...863L..18A, 2018MNRAS.481.2711G, 2018PhRvL.120x1103L, 2018Natur.561..355M, 2019ApJ...883L...1F, 2019ApJ...886L..17H, 2019ApJ...870L..15L, 2019ApJ...880L..23W}. In this case, the radiation from the energetic core is subdominant at early times because it is beamed away due to ultrarelativistic motion. On the other hand, the less energetic tail close to the LOS becomes important in the modeling of early emission. As the jet decelerates, the LOS eventually enters the beaming cone and the energetic core gradually becomes visible. 
This process leads to a slowly increasing luminosity in the early phase of the afterglow, which effectively explains the rising light curves observed in the GRB 170817A afterglow. In contrast to a structured jet, a top-hat jet model predicts a sharp increase in the early phase for a misaligned jet, which is at odds with observations from this event (e.g., \citealt{2018MNRAS.478..407N}). Additionally, the apparent superluminal motion observed in this event \citep{2018Natur.561..355M, 2019Sci...363..968G, 2022Natur.610..273M} further supports an off-axis observation of a structured jet. It has been suggested that future multimessenger observations of neutron star mergers will predominantly be off-axis (e.g., \citealt{2017Natur.551...71T, 2018MNRAS.473L.121K,2018MNRAS.474.5340S,2019A&A...631A..39D,2019MNRAS.483..840B,2021ApJ...908..200W}), indicating that the GW170817 and GRB 170817A scenario is likely to be common.

Various works have suggested that the rising slope of off-axis GRB afterglows helps to constrain the observing angle between the jet axis and the LOS (e.g., \citealt{2018ApJ...860L...2F, 2018MNRAS.478.4128G, 2019ApJ...880L..23W, 2020ApJ...896..166R}). Such a constraint becomes particularly important when considering the combined view with gravitational wave observations. The observing angle, which is also assumed to be the inclination angle of the premerger neutron star binary, is degenerate with the luminosity distance inferred from gravitational wave data. A better constrained observing angle helps break the degeneracy and provides a better distance measurement, making it a powerful tool in the standard siren study of the Hubble constant (e.g., \citealt{2017ApJ...851L..36G, 2019NatAs...3..940H, 2021ApJ...908..200W, 2023ApJ...943...13W, 2024MNRAS.528.2600G}). This is particularly helpful in the current context, where there is growing evidence of tension between different measurements of the Hubble constant (see \citealt{2021CQGra..38o3001D} for a review). So far, all studies utilizing this method have been model-dependent because they assume a specific type of jet profile. The jet profile is either approximated by an analytical expression, such as Gaussian jets (e.g., \citealt{2018MNRAS.478.4128G, 2020ApJ...896..166R}), power-law jets (e.g., \citealt{2022Natur.610..273M}), and boosted fireball jets (e.g., \citealt{2018ApJ...869...55W, 2019ApJ...880L..23W}), or adopted by the output of another numerical simulation (e.g., \citealt{2021ApJ...908..200W}). It is unclear how the assumed jet profile affects the inferred observing angle. Recently, \citet{2024MNRAS.528.2600G} found a moderate difference in the observing angle inferred from Gaussian and power-law jets, which motivates further investigation into the systematic error caused by jet structures and reconstructs the jet structure from the observation \citep{2020MNRAS.497.1217T}. Therefore, it is necessary to develop a reliable afterglow model that allows for the flexibility of various jet structures.

Modeling the GRB afterglow with an arbitrary jet structure presents challenges. Important features of off-axis afterglow can be well modeled by analytic methods \citep{2020MNRAS.493.3521B,2022MNRAS.515..555B}, but the accurate modeling of late-time spreading effect \citep{1999ApJ...525..737R, 1999ApJ...519L..17S} remains a challenge. This complexity stems from the fact that spreading, driven by internal pressure, is a nonlinear hydrodynamic process. Consequently, analytical methods struggle to accurately describe this phase. Several attempts, including approximations based on sound speed expansion (e.g., \citealt{1999ApJ...525..737R, 2020ApJ...896..166R, 2023MNRAS.520.2727N}) or curvature effects (e.g., \citealt{2012MNRAS.421..570G}), have been made to model the spreading of a top-hat jet. Some of them are even calibrated to numerical simulations (e.g., \citealt{2018ApJ...865...94D}). However, it remains uncertain to what extent these semianalytical methods apply to structured jet models. To date, the most accurate and reliable approach for modeling in these scenarios is relativistic numerical hydrodynamic simulations. 

Fully numerical hydrodynamic approaches also have challenges. In the ultrarelativistic limit, the swept-up materials are compressed in an extremely thin shell, typically with a width of approximately $R/\gamma^2$ where $R$ represents the shock radius and $\gamma$ represents the Lorentz factor. Numerically resolving this thin structure is challenging, especially when the jet's Lorentz factor is exceptionally high. This substantial computational cost restricts the use of Markov Chain Monte Carlo (MCMC) for parameter estimation in models with varying jet structures. Consequently, there is a pressing need to develop a model that balances the computational efficiency of semianalytic methods with the accuracy of full numerical hydrodynamics.

In this work, we propose an alternative hydrodynamic approach that significantly reduces the computational cost of a full hydrodynamic simulation. The concept of our approach is similar to that of \citet{2003ApJ...591.1075K} and \citet{2020arXiv200510313L}, where the shocked region is approximated by an infinitely thin two-dimensional surface. By additionally assuming the axial symmetry of the jet, the simulation is effectively simplified into a one-dimensional representation. While our method maintains the accuracy of numerical hydrodynamics, it attains a computational efficiency that is comparable to semianalytic methods. We have developed the methods described in this work into a numerical tool \texttt{jetsimpy}. This code simulates the evolution of the blast wave and generates synthetic afterglow observables for a jet with tabulated angular energy and Lorentz factor profiles, traveling within either an interstellar medium (ISM) environment or stellar wind environment. 

This paper is organized as follows. In \S \ref{sec:numeric} we describe our hydrodynamic model of a structured GRB jet. In \S \ref{sec:hydro} we present our numerical results and the comparisons to full relativistic hydrodynamics simulation. In \S \ref{sec:afterglow} we briefly introduce our modeling of the afterglow observables. In \S \ref{sec:test} we verify and compare our methods to existing afterglow modeling tools. In Section \ref{sec:observation}, we apply our model to the case of GRB 170817A. Finally, in \S \ref{sec:summary}, we discuss and summarize our methods.

\section{Hydrodynamic Model} \label{sec:numeric}

\subsection{Basic assumptions} \label{subsec:assumption}
In this work, we assume the jet is axial-symmetric and the external medium follows a power-law density distribution: $\rho_0\propto r^{-k}$. When the ultrarelativistic jet moves through a cold external medium, it gives rise to a forward-reverse shock structure. The region between these two shocks is commonly referred to as the ``blast". The forward shock sweeps up the external material, increasing the mass in the blast, while the reverse shock propagates through the ejecta in the opposite direction. In this study, we focus on the forward shock and assume that the ejecta is fully shocked and has sufficiently cooled down. That is, we consider the blast region to consist of a hot, swept-up material layer and a cold ejecta layer. We further ignore the impact of radiative loss on the hydrodynamics. Throughout this section, variables associated with the ejecta are labeled as ``ej" and those related to the swept-up material are labeled as ``sw".

To describe the thermodynamic properties of the fluid, we consider two reference frames: the comoving frame, which moves along with the fluid elements, and the burster frame, which remains static relative to the burst origin. All physical quantities measured in the comoving frame are indicated with a prime ('). On the other hand, quantities in the burster frame are denoted without the prime. The velocity $\beta$ and Lorentz factor $\gamma$ are always measured in the burster frame.

The thermodynamic properties of the fluid elements just behind the forward shock can be determined by the shock jump condition \citep{2011ApJ...733...86U}:
\begin{align}
    \rho'_{\rm sw} &= 4 \gamma \rho_0 \label{eq:density} \\
    e'_{\rm sw} &= 4 \gamma(\gamma - 1) \rho_0 c^2 \label{eq:energy_density} \\
    p'_{\rm sw} &= \frac{4}{3}\beta^2\gamma^2\rho_0 c^2 \label{eq:pressure}
\end{align}
In the last expressions, $\rho'_{\rm sw}$, $e'_{\rm sw}$, and $p'_{\rm sw}$ represent the mass density, energy density (rest mass excluded), and pressure of the shocked material in the comoving frame, respectively. The $\rho_0$ denotes the mass density of the ambient matter. These relations are derived from a semianalytic trans-relativistic equation of state, which smoothly interpolates the adiabatic index between the ultrarelativistic regime and Newtonian regime \citep{2007MNRAS.378.1118M}.

In our model, the blast shell is approximated as an infinitely thin, two-dimensional surface. Within this approximation, the thermodynamic properties of both the hot, swept-up material and the cold ejecta are effectively represented by Dirac delta functions. We further assume that the radial average of the swept-up material is proportional to the post-shock values. This allows us to directly determine the scale of the delta function based on the shock jump conditions. Our approximations remain valid due to the following considerations:

(i) {\it The blast shell is effectively thin.} We can estimate the typical width of the blast shell, $\Delta R$, by noting that the total mass behind the forward shock equals the total swept-up mass at radius $R$. For example, in a constant external medium, it indicates that $4\pi R^2 \Delta R\gamma\rho'_{\rm sw}\approx 4\pi R^3\rho_0/3$, yielding an estimate of $\Delta R \sim R/12\gamma^2$ in the burster frame. This estimation indicates that a thin shell is an excellent approximation in the ultrarelativistic limit, hence motivating the delta function approximation. While this assumption might not hold as well in the Newtonian regime, the use of a trans-relativistic equation of state ensures that the radially integrated mass and energy scale correctly with the well-known Sedov-Taylor solution (e.g., \citealt{1959sdmm.book.....S}). As a result, our model continues to provide order-of-magnitude estimations in this limit.

(ii) {\it The radial profile peaks at the shock.} While the radial profile of the blast may evolve with time, it is, in general, a narrow function that peaks at the shock, especially in the relativistic regime. This means that the post-shock value is a good first-order estimation of the radial average, so that we can approximate the blast by a homogeneous shell. The scale of the delta function can thereby determined by the shock jump condition. It is also possible to calibrate the scale with self-similar solutions. 

Based on these considerations, we first define the following radial integrated values: the mass of swept-up matter and ejecta {\it per solid angle}
\begin{align}
    M_{\rm sw} & = \int \gamma\rho'_{\rm sw}r^2dr, \\
    M_{\rm ej} & = \int \gamma\rho'_{\rm ej}r^2dr,
\end{align}
where $\rho'_{\rm ej}$ is the density of the ejecta. Within the infinitely thin shell approximation, the densities are assumed to be delta functions. In this limit, the above integral implies the following relations
\begin{align}
    \rho'_{\rm sw} & = M_{\rm sw}R^{-2}\gamma^{-1}\delta(r-R) \label{eq:rho_sw}\\
    \rho'_{\rm ej} & = M_{\rm ej}R^{-2}\gamma^{-1}\delta(r-R) \label{eq:rho_ej}
\end{align}
where $\delta(x)$ is the Dirac delta function. Similarly, we can approximate the energy density and pressure using delta functions. According to the shock jump condition, the density just behind the shock is $\rho'_{\rm sw} = 4 \gamma \rho_0$. We can construct delta functions for energy and pressure by substituting all terms of ``$4\gamma\rho_0$" in the shock jump conditions with the above delta function. Given that the radial profile of the blast is self-similar, the scale is proportional to this value with a calibration coefficient. These considerations lead to the following approximations:
\begin{align}
    \bar{e}'_{\rm sw} & = s(1-\gamma^{-1})M_{\rm sw}R^{-2}c^2\delta(r-R), \label{eq:e_sw}\\
    \bar{p}'_{\rm sw} & = \frac{1}{3}s\beta^2M_{\rm sw}R^{-2}c^2\delta(r-R). \label{eq:p_sw}
\end{align}
Our calibration coefficient ``$s$" will be determined by matching our solutions to analytic and numerical results later. These relations will serve as the building blocks in the derivation of evolution equations.

\subsection{Evolution equations}
To derive the evolution equations, we start from the relativistic Euler equations,
\begin{align}
    & T^{\mu\nu}_{;\mu}=0, \\
    & J^{\mu}_{;\mu} = 0,
\end{align}
where $T^{\mu\nu}$ is the energy-momentum tensor and $J^{\mu}$ is the mass current. These relations represent the conservation of energy, momentum, and mass. In this work, we assume that the blast is perfect fluid. In a spherical coordinate with metric tensor $g_{\mu\nu}={\rm diag}(-1, 1, r^2, r^2\sin^2\theta)$, the energy-momentum tensor and mass current take the following form
\begin{align}
    & T^{\mu\nu} = h'_{\rm b}u^{\mu}u^{\nu} + g^{\mu\nu}\bar{p}'_{\rm sw}, \\
    & J_{\rm sw}^{\mu} = \rho'_{\rm sw}u^{\mu}, \\
    & J_{\rm ej}^{\mu} = \rho'_{\rm ej}u^{\mu},
\end{align}
where $h'_{\rm b}$ is the enthalpy of the blast defined by
\begin{equation}
    h'_{\rm b} = \bar{e}'_{\rm sw} + \bar{p}'_{\rm sw} + \rho'_{\rm sw}c^2 + \rho'_{\rm ej}c^2.
\end{equation}
Assuming axial symmetry, the 4-velocity $u^{\mu}$ is
\begin{equation}
    u^{\mu} = (\gamma, \beta_{\rm r}\gamma, \frac{\beta_{\theta}\gamma}{r}, 0)
\end{equation}
where $\beta_{\rm r}$ and $\beta_{\theta}$ are the velocity components along radial and polar directions, respectively. In the definitions above, the swept-up mass and ejecta in the mass current are defined independently, while the energy-momentum tensor is defined by the sum. This is because energy conservation applies to the entire blast, whereas mass conservation holds for each component individually. By evolving the total energy-momentum tensor, we ensure the conservation of total energy, while avoiding the details of the interaction between ejecta and swept-up material.

In the spherical coordinate, the Eulerian equations can be explicitly expressed by the following vector form
\begin{equation}\label{eq:eulerian}
    \frac{1}{c}\frac{\partial {\bf A_{\rm t}}}{\partial t} + \frac{1}{r^2}\frac{\partial (r^2{\bf A_{\rm r}})}{\partial r} + \frac{1}{r\sin\theta}\frac{\partial ({\bf A_{\theta}}\sin\theta)}{\partial\theta} + {\bf s} =0,
\end{equation}
where the vector terms are defined by
\begin{gather*}
    {\bf A_{\rm t}} = 
    \begin{pmatrix}
        h'_{\rm b}\gamma^2 - p'_{\rm sw} \\
        h'_{\rm b}\gamma^2\beta_{\theta} \\
        \rho'_{\rm sw}\gamma \\
        \rho'_{\rm ej}\gamma 
    \end{pmatrix},\ 
    {\bf A_{\rm r}} = 
    \begin{pmatrix}
        h'_{\rm b}\gamma^2\beta_{\rm r} \\
        h'_{\rm b}\gamma^2\beta_{\theta}\beta_{\rm r} \\
        \rho'_{\rm sw}\gamma\beta_{\rm r} \\
        \rho'_{\rm ej}\gamma\beta_{\rm r}
    \end{pmatrix},
\end{gather*}
\begin{gather*}
    {\bf A_{\theta}} = 
    \begin{pmatrix}
        h'_{\rm b}\gamma^2\beta_{\theta} \\
        h'_{\rm b}\gamma^2\beta_{\theta}^2+p'_{\rm sw} \\
        \rho'_{\rm sw}\gamma\beta_{\theta} \\
        \rho'_{\rm ej}\gamma\beta_{\theta} 
    \end{pmatrix},\ 
    {\bf s} = \frac{1}{r}
    \begin{pmatrix}
        0 \\
        h'_{\rm b}\gamma^2\beta_{\theta}\beta_{\rm r} - \frac{\cos\theta}{\sin\theta}p'_{\rm sw} \\
        0 \\
        0
    \end{pmatrix}.
\end{gather*}
In these equations, the {\it r}-component momentum equation is not included. The explanation will follow after we derive the radial integrated equations.

Now, we can apply the delta function approximation of the thermodynamic quantities to eq. \ref{eq:eulerian}. After this, we multiply this equation by $r^2$ and integrate it over the blast shell radius. For the sake of convenience, we define the following integrated variables
\begin{align}
    E_{\rm b} & = \frac{1}{c^2}\int_0^R (h'_{\rm b}\gamma^2-\bar{p}'_{\rm sw})r^2dr \nonumber \\
    & = s(1 + \frac{1}{3}\beta^4)\gamma^2M_{\rm sw} + (1-s)\gamma M_{\rm sw} + \gamma M_{\rm ej}, \label{eq:energy_relation} \\
    P_{\rm sw} & = \frac{1}{c^2}\int_0^R \bar{p}'_{\rm sw}r^2dr = \frac{1}{3}s\beta^2 M_{\rm sw}, \label{eq:pressure_relation}\\
    H_{\rm b} & = \frac{1}{c^2}\int_0^R h'_{\rm b}\gamma^2r^2dr = E_{\rm b} + P_{\rm sw}.\label{eq:enthalpy_relation}
\end{align}
These variables represent the radial integrated energy, pressure, and enthalpy, each normalized to the unit of mass by dividing them by $c^2$. Utilizing these variables, we can express the radial integrated Eulerian equations as follows:
\begin{equation}\label{eq:evolution_equation}
    \frac{\partial {\bf U}}{\partial t} + \frac{1}{\sin\theta}\frac{\partial ({\bf F}\sin\theta)}{\partial \theta} + {\bf S} = 0
\end{equation}
where the conserved variables ${\bf U}$, flux vector ${\bf F}$, and source terms $\bf S$ are defined by
\begin{gather}
    {\bf U} = 
    \begin{pmatrix}
        E_{\rm b} \\
        \beta_{\theta}H_{\rm b} \\
        M_{\rm sw} \\
        M_{\rm ej} 
    \end{pmatrix}, \\
    {\bf F} = \frac{c}{R}
    \begin{pmatrix}
        \beta_{\theta}H_{\rm b} \\
        \beta_{\theta}^2 H_{\rm b} + P_{\rm sw} \\
        \beta_{\theta} M_{\rm sw} \\
        \beta_{\theta} M_{\rm ej}
    \end{pmatrix}, \\
    {\bf S} = \frac{c}{R}
    \begin{pmatrix}
        - \frac{1}{c}\frac{\partial R}{\partial t}\rho_0 R^3 \\
        \beta_{\theta}\beta_{\rm r}H_{\rm b} - \frac{\cos\theta}{\sin\theta}P_{\rm sw} \\
        - \frac{1}{c}\frac{\partial R}{\partial t}\rho_0 R^3 \\
        0
    \end{pmatrix}.
\end{gather}
In these equations, additional source terms emerge in both the energy and mass equations. These terms result from the movement of the integral boundary, specifically the forward shock. They account for the accumulation of external medium by the shock.

To complete the set of equations, we also need the evolution of the blast shell radius $R(t, \theta)$. This can be derived by considering the speed of a fluid element, given by $dR/dt=\beta c$. It is important to note that in the infinitely thin shell approximation, the velocity ($\beta$) difference between the blast and the forward shock is essentially disregarded. While this is a reasonable approximation in the ultrarelativistic limit, it becomes less valid in the Newtonian limit. To address this discrepancy, we substitute the velocity of the blast with the velocity of the forward shock:
\begin{equation}\label{eq:shock_velocity}
    \beta_{\rm f} = \frac{4\beta\gamma^2}{4\gamma^2-1}
\end{equation}
This can be better understood by defining the radius of the blast as the position of the forward shock. Now we can derive the radius evolution equation by changing from a total derivative to a local derivative:
\begin{equation}\label{eq:radial}
    \frac{\partial R}{\partial t} = \beta_{\rm f}c - \frac{\partial R}{\partial \theta}\frac{\beta_{\theta}c}{R}
\end{equation}
In the numerical tests, we find that replacing $\beta$ by $\beta_{\rm f}$ in this equation improves the accuracy of the solutions. 

Now, the evolution of the blast wave can be solved by coupling eq.\ref{eq:evolution_equation} with eq.\ref{eq:radial}, provided with a set of appropriate initial conditions. These initial conditions could be any physically reasonable angular distribution of initial energy $E_{\rm b,0}$ and Lorentz factor $\gamma_0$. The mass of the ejecta can then be estimated by $M_{\rm ej,0}=E_{\rm b,0}/\gamma_0$. 

Similar to original Euler equations, these equations are nonlinear and require a well-designed approach to solve and maintain numerical stability. The details of the numerical scheme are explained in Appendix \ref{appendix:scheme}.

The approach described in this section is similar to previous works \citep{2003ApJ...591.1075K, 2020arXiv200510313L} given that they all simplify the hydrodynamic equations by infinitely thin shell approximation. However, there are several major differences between these approaches.

The major difference between our work and \citealt{2003ApJ...591.1075K} is that the radial momentum equation is not included in our work. This is because the energy, momentum, and mass in \citealt{2003ApJ...591.1075K} are solved by a complete set of hydrodynamic equations, while in our approach the energy and mass are related by shock jump condition. The direct incorporation of the shock jump condition provides extra information. As a result, the velocity can be directly solved by energy conservation (eq. \ref{eq:energy_relation}), which does not need momentum terms. This means that we must exclude one of the momentum equations from the original equation set to maintain self-consistency. We propose to exclude the radial momentum because the polar momentum is necessary to determine the tangent velocity of a fluid element. The removal of one equation increases the computational efficiency. It also helps to reduce the order of the Jacobian matrix and simplify its eigenstructure, which is crucial in constructing the numerical scheme (see Appendix \ref{appendix:scheme}). In addition, eq. \ref{eq:energy_relation} offers a convenient way of calibrating the model to self-similar solutions (see the next section).

The other similar approach introduced in \citealt{2020arXiv200510313L} also directly incorporates the shock jump condition to approximate the energy and density in a way that is similar to ours. In their method, the evolution of a fluid element is solved by a Lagrangian approach, where the lateral motion of a grid point on the surface is determined by the total surrounding pressure. This is fundamentally different from our Eulerian approach. 
While the Lagrangian approach is more suited to complex flow motion, the Eulerian approach is better at achieving a higher order of accuracy and capturing the nonlinear waves. The comparison between the two approaches in modeling the jet spreading needs further investigation.

We notice that our hydrodynamic model introduces minor inconsistencies. Specifically, the direction predicted for a fluid element by our model does not always align with the shock normal direction. This inconsistency originates from the infinitely thin shell approximation, which is insufficient to reproduce all predictions of a full hydrodynamic model. On the other hand, assuming that fluid elements move along the shock normal direction implies that the expansion of a thin shell is completely determined by the geometry of the shock front, instead of being governed by hydrodynamic equations. By accepting this minor inconsistency, our approach allows us to derive the evolution equations from first principles (i.e., the Euler equations), which should provide better accuracy than a purely geometric approach.

\subsection{Calibration with self-similar solutions}
The calibration coefficient $s$ appearing in the equations is determined based on comparisons with exact self-similar solutions. It is well-known that a blast wave follows the Blandford-McKee self-similar solution in the ultrarelativistic limit \citep{1976PhFl...19.1130B}, and follows the Sedov-Taylor solution in the Newtonian limit (e.g., \citealt{1959sdmm.book.....S}). Calibration can be achieved by aligning eq. \ref{eq:energy_relation} with the radial integral of these self-similar solutions. It is also possible to calibrate the coefficient in both limits and smoothly vary it in the transitioning phase.

To calibrate in the relativistic limit, we first revisit the Blandford-McKee solutions \citep{1976PhFl...19.1130B}. In a power-law external density medium described by $\rho_0(r)= Ar^{-k}$, the pressure and Lorentz factor profiles are as follows:
\begin{align}
    & p'_{\rm BM}(r) = \frac{2}{3}\rho_0 c^2 \Gamma_{\rm f}^2f(\chi), \\
    & \gamma_{\rm BM}(r)^2 = \frac{1}{2} \Gamma_{\rm f}^2 g(\chi),
\end{align}
where the self-similar variable and functions are defined by
\begin{align}
    & f(\chi) = \chi^{-(17-4k)/(12-3k)}, \\
    & g(\chi) = \chi^{-1}, \\
    & \chi = 1 + 2(4 - k)(1-\frac{r}{R})\Gamma_{\rm f}^2.
\end{align}
In these equations, $\Gamma_{\rm f}=\sqrt{2}\gamma_{\rm BM}(R)$ represents the Lorentz factor of the forward shock\footnote{The Lorentz factor $\Gamma_{\rm f}$ does not precisely match with the one corresponding to the aforementioned velocity $\beta_{\rm f}$ (eq. \ref{eq:shock_velocity}), because the equation of state used in the Blandford-McKee solution is not trans-relativistic.}. Based on the provided profile, the energy per solid angle inside the forward shock can be integrated as follows:
\begin{equation}\label{eq:bm_energy}
    E_{\rm b, BM} = \int_0^R 4p'_{\rm BM}\gamma_{\rm BM}^2 r^2dr.
\end{equation}
The mass per solid angle can be simply calculated by integrating the external density profile within the radius $R$:
\begin{equation}
    M_{\rm BM} = \frac{A}{3-k}R^{3-k}.
\end{equation}

To compare these results to eq. \ref{eq:energy_relation}, we first define the rest mass and ejecta mass excluded energy
\begin{equation}
    E_{\rm sw}=E_{\rm b}-\gamma M_{\rm ej}-M_{\rm sw}.
\end{equation}
In the ultrarelativistic limit, this value approaches $E_{\rm sw}=\frac{4}{3}s\gamma^2 M_{\rm sw}$. In the same limit, eq. \ref{eq:bm_energy} approaches $E_{\rm b, BM}\propto \gamma^2 M_{\rm BM}$. To align the two values, we define the calibration coefficient in this limit by
\begin{equation}
    s_{\rm BM}(k) = \lim_{\gamma\to\infty} \frac{3E_{\rm b, BM}}{4\gamma^2 M_{\rm BM}}.
\end{equation}
We find that this coefficient depends on $k$ only. 

\begin{figure}[ht!]
\centering
\includegraphics[width=\columnwidth]{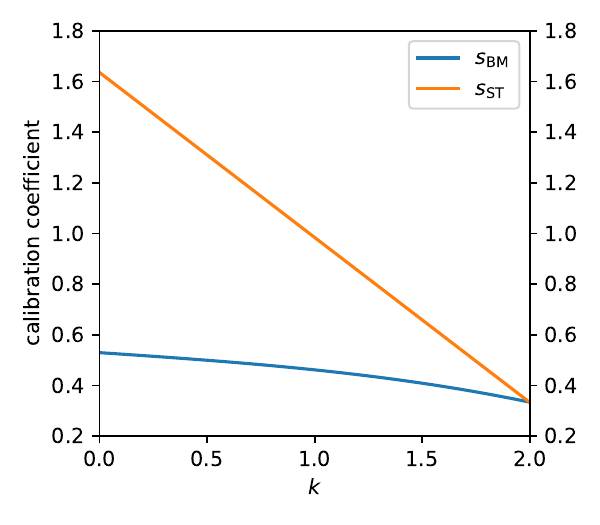}
\caption{The coefficients to calibrate our method with Blandford-McKee and Sedov-Taylor self-similar solutions.}
\label{fig:calib}
\end{figure}

We also calibrate our model with the Sedov-Taylor solution in a similar manner. Since the exact solution is nontrivial, we numerically calculate the radial integrated energy $E_{\rm ST}$ using the publicly available tool \texttt{sedov.tbz} \citep{Kamm2007}\footnote{\url{https://cococubed.com/research_pages/sedov.shtml}}. The mass integral remains the same as above, i.e., $M_{\rm ST}=M_{\rm BM}$. In the Newtonian limit, our model approaches $E_{\rm sw}= \frac{1}{2}(1+s)\beta^2 M_{\rm sw}$, while the Sedov solution approaches $E_{\rm b, ST}\propto \beta^2M_{\rm ST}$. As a result, the calibration coefficient in this limit can be defined by
\begin{equation}
    s_{\rm ST}(k) = \lim_{\beta\to 0} \frac{2E_{\rm b, ST}}{\beta^2 M_{\rm ST}}-1.
\end{equation}
We find the coefficient also depends on $k$ only. Both calibration coefficients are nontrivial functions of $k$, and they are presented in Fig. \ref{fig:calib}.

It can be seen that the two coefficients are in general different. To smoothly transition between the two limits, we propose the following approximation for the total calibration coefficient:
\begin{equation}
    s(k)=\frac{s_{\rm ST} + 2s_{\rm BM}\beta^2\gamma^2}{1 + 2\beta^2\gamma^2}.
\end{equation}
The factor 2 is chosen to better agree with hydrodynamics results in respective limits.

We can further extend this result to apply in a more sophisticated external medium by defining a ``local" power index for the radial profile: $k=-d\log\rho_0/d\log r$. We assume that the evolution of the calibration coefficient in an environment with varying $k$ is quasi-static. This approach is particularly useful in scenarios where the external medium transitions smoothly from a wind-like to an ISM-like environment.

We caution here that the calibration in the Newtonian limit is tentative because the thin shell approximation will eventually break down at very low speed. In this regime, the extensive radial profile leads to hydrodynamic and radiation processes that differ significantly from those in the relativistic regime. Therefore, our modeling in the Newtonian limit should be regarded as providing only an order-of-magnitude estimation.

\subsection{\texttt{jetsimpy}}
The hydrodynamic model introduced above, along with the modeling of the synchrotron emission from the electrons accelerated at the forward shock to be discussed in subsequent sections, have been constructed into a numerical tool, \texttt{jetsimpy}, and is provided to the community \citep{wang_2024_11078597}\footnote{\url{https://github.com/haowang-astro/jetsimpy}}.  The code is written in C++ and features a flexible and user-friendly Python interface. It is designed for arbitrary initial energy and Lorentz factor distribution as initial conditions and can evolve the jet in ISM-like, wind-like, or mixed profiles. \texttt{jetsimpy} drastically reduces the computational cost by a factor of thousands compared to numerical simulation. This efficiency is particularly beneficial for computing millions of cases with varying angular profiles, such as in an MCMC study. Furthermore, as we will demonstrate in the following sections, our tool maintains a level of accuracy comparable to that of full numerical simulations, even in situations where nonlinear hydrodynamic effects are significant.

\section{Hydrodynamic Results}\label{sec:hydro}
In this section, we employ our model to simulate the evolution of a jet and compare the results with those obtained from full numerical hydrodynamics. We test our solutions for a top-hat jet moving in two different mediums: a constant interstellar medium (ISM), where $k=0$, and a wind medium, where $k=2$. Compared to structured jets, a top-hat jet exhibits more significant nonlinear hydrodynamic effects due to its sharp edge, such as the emergence of a bow shock around the jet. This feature presents a challenging scenario to test our code under extreme conditions.

The full numerical hydrodynamics simulations are conducted using the moving-mesh code \texttt{JET} \citep{2013ApJ...775...87D}. In this code the grid lines move along the radial directions at local speeds of fluid, thereby capable of capturing flow structures with high resolution and high Lorentz factors. The simulations are two-dimensional and are performed in spherical coordinates, assuming axial symmetry.

\subsection{The jet evolution}
\label{sec:jet_evolution}
To begin with, we present the typical behavior of jet evolution as predicted by our model. A jet generally experiences four stages during its evolution (e.g., \citealt{1999ApJ...513..669K}; see \citealt{2015PhR...561....1K} for a review and references therein). Initially, the total mass of the blast shell is dominated by the ejecta carried with the jet, and the jet coasts at a constant velocity. After the accumulated mass exceeds the mass of the ejecta, the jet begins to decelerate, following the Blandford-McKee solution. When the jet becomes sub-relativistic, the fluid elements along the surface of the blast wave become causally connected, leading to the jet starting its lateral spreading due to internal pressure. Eventually, the jet transitions to a Newtonian phase and the shape of the blast wave becomes mostly spherical. The subsequent evolution is best described by the Sedov-Taylor solution.

\begin{figure}[ht!]
\centering
\includegraphics[width=\columnwidth]{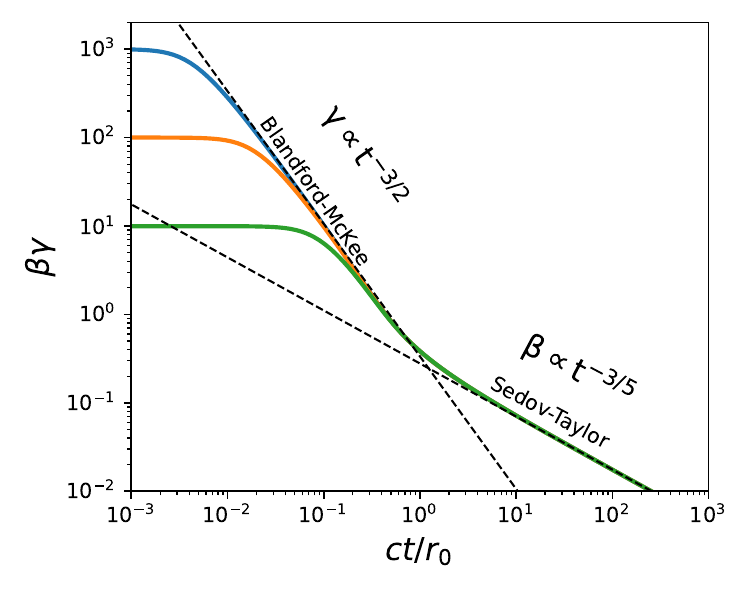}
\caption{The evolution of proper velocity for isotropic explosion with various initial Lorentz factors ($\gamma_0$=10, 100, and 1000 for green, orange, and blue lines, respectively). The time is scaled by $r_0/c$ which is the time when the jet becomes sub-relativistic. The dashed lines are the exact Blandford-McKee and Sedov-Taylor self-similar solutions.
\label{fig:bm_st}}
\end{figure}

In Fig. \ref{fig:bm_st} we show the evolution of proper velocity for isotropic blast waves with varying initial Lorentz factors moving in an ISM environment. The time is normalized by a scale $r_0/c$, where $r_0=(3E_{\rm iso}/4\pi\rho_0 c^2)^{1/3}$ is the radius at which the jet becomes sub-relativistic. For comparison, the exact self-similar solutions are also presented. Our results align with both the exact Blandford-McKee and Sedov-Taylor solutions in their respective limits and exhibit a smooth transition in between.

\begin{figure*}[!t]
\centering
    \includegraphics[width=\textwidth]{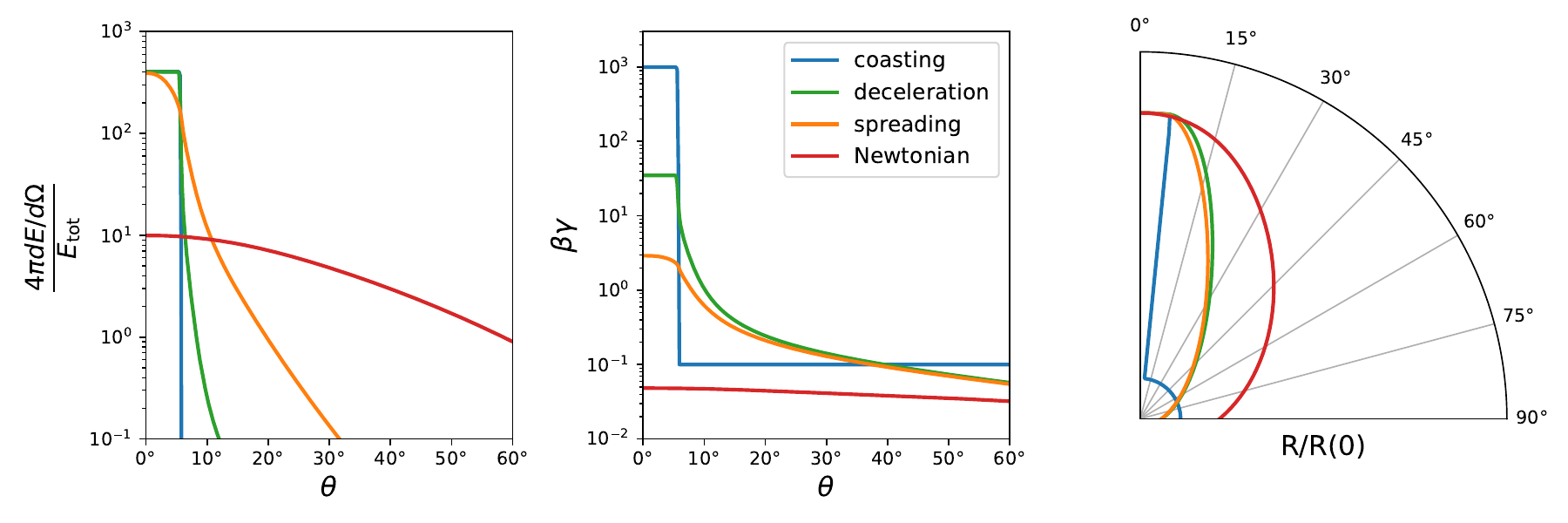}
\caption{The evolution of a top-hat jet with a half-opening angle of 0.1 radians traveling in a constant density interstellar medium. Different phases of the jet are represented by the corresponding colors: blue lines the coasting phase, green lines the deceleration phase, orange lines the spreading phase, and red lines the Newtonian phase. Left-hand panel: normalized energy. Middle panel: proper velocity. Right-hand panel: radius scaled with $R(\theta=0)$.
\label{fig:jet_evolution}}
\end{figure*}

In Figure \ref{fig:jet_evolution}, we show the evolution of angular structure for a top-hat jet with a half-opening angle of 0.1 radians moving through ISM. The four colors in the figure correspond to the snapshots of four phases of the jet's evolution, as indicated in the labels. During the deceleration phase (green lines), a bow shock emerges at the edge of the jet, forming a sub-relativistic tail. As the core decelerates to mildly relativistic speeds (orange lines), the jet begins lateral expansion due to its internal pressure. In the Newtonian phase (red line), the jet gradually becomes spherical.

\subsection{Comparison with full numerical hydrodynamics}
\label{sec:compare_hydro}
We further compare our model to full numerical hydrodynamic code \texttt{JET}, in particular the evolution of jet structure during the spreading phase. 

\begin{figure}[ht!]
\centering
    \includegraphics[width=\columnwidth]{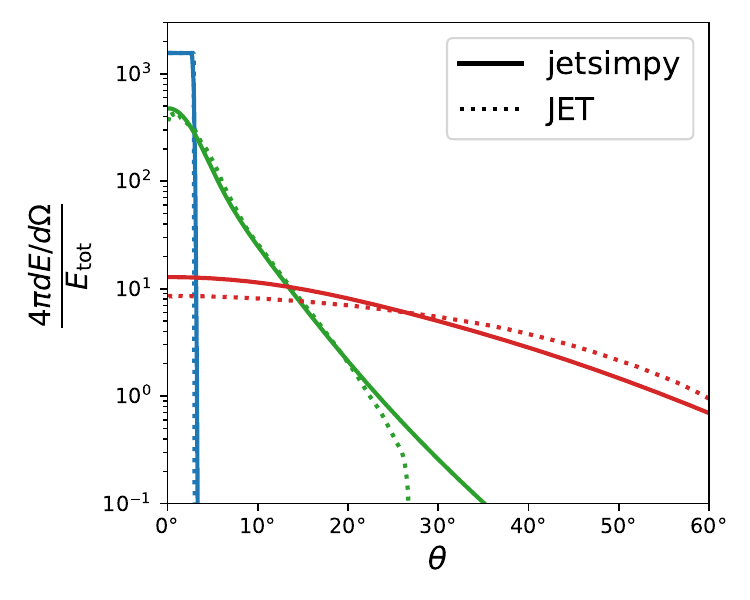}
\caption{A comparison between \texttt{jetsimpy} and the numerical hydrodynamic code \texttt{JET}. The numerical setup is a top-hat jet with a half-opening angle of 0.052 radians (3 degrees) traveling in a uniform external medium. We present a comparison of three stages during the evolution: the blue lines that represent the initial condition, the green lines that represent the intermediate stage of spreading, and the red lines that represent the Newtonian phase.
\label{fig:compare_numeric_ism}}
\end{figure}

In the first case, we consider a top-hat jet with a half-opening angle of 0.052 radians (3°) traveling through ISM. The initial condition for the \texttt{JET} simulation is a Blandford-McKee radial profile. In Fig. \ref{fig:compare_numeric_ism} we show the evolution of normalized energy (rest mass excluded) at three stages: the initial condition (blue lines), the spreading phase (green lines), and the Newtonian phase (red lines). The comparison reveals good agreement with the core of the jet and shows a more extended low-energy tail outside the core.

\begin{figure}[ht!]
\centering
    \includegraphics[width=\columnwidth]{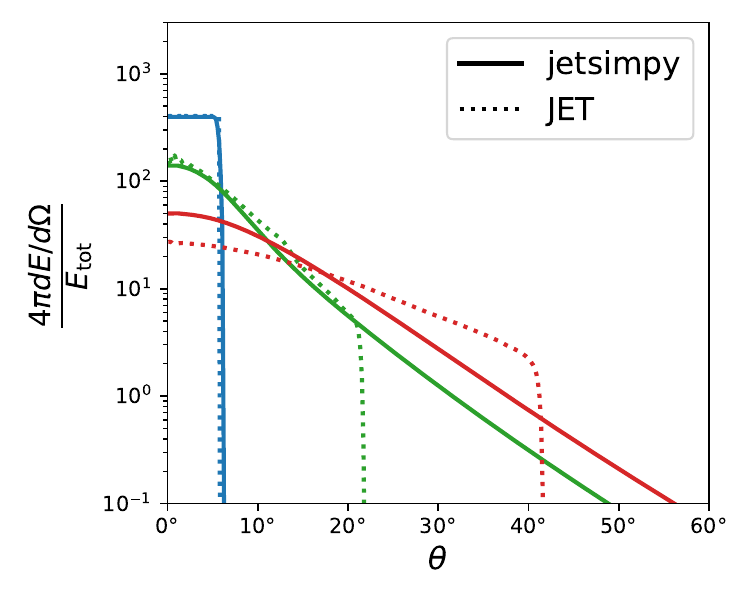}
\caption{Same as the last figure, but for a top-hat jet with a half-opening angle of 0.1 radians (6°) traveling in a wind medium.
\label{fig:compare_numeric_wind}}
\end{figure}

In the second comparison, we consider a top-hat jet with a half-opening angle of 0.1 radians (6°) traveling in a wind medium. The result is shown in Fig. \ref{fig:compare_numeric_wind}. This comparison also reveals overall good agreement but is somewhat less accurate than in the ISM case. {The slightly increased discrepancy in the wind medium test may indicate that our model is less accurate if the density of the external medium varies steeply as a function of radius.} Consequently, we consider our model more accurate in the ISM environment as compared to the wind medium. 

The major discrepancy appearing in the two tests is that our model generates a more extended tail at large angles. This discrepancy arises from the infinitely thin shell approximation, where the entire radial profile is projected onto a single surface. As a result, the regions inside and outside of the edge of the jet become adjacent to each other, although they are in fact spatially distant along the radial direction. This effect leads to energy exchange between the regions that do not occur in practice. Nevertheless, In our tests, we find this effect is practically unimportant in the afterglow modeling, even if the observer's LOS lies within the region of the extended tail. We notice that the condition for producing the extended tail is a significant radial velocity gradient, which only appears in the Newtonian regime. As a result, the radiation from this tail is not beamed, and the dominant source of radiation is always from the well-modeled region.

\subsection{Computational performance}
In addition to the good agreement, our code performs significantly faster than full numerical hydrodynamics. The performance boost can be attributed to two aspects. First, with one less dimension one can significantly reduce the required number of grid cells of the simulation. Second, the Courant–Friedrichs–Lewy (CFL) condition (see Appendix \ref{appendix:scheme}) is more relaxed than a full simulation. The CFL condition limits the simulation time step to the signal-crossing time of the grid cells to maintain numerical stability. For a moving-mesh code such as \texttt{JET}, the signal speed is effectively the sound speed along two directions: the radial direction and the polar direction. In general, the time step is limited by the radial sound-crossing time. However, in our approach, the simulation is performed only along the polar direction, so the time step is only limited by the angular sound-crossing time. As the blast wave expands, the angular speed decreases due to the increasing radius. Consequently, the time step can grow progressively larger as the jet evolves, leading to an improved computational efficiency.

\section{Afterglow modeling}\label{sec:afterglow}
In this section, we describe the modeling of GRB afterglows by post-processing the numerical data obtained from our hydrodynamic models. In the current stage of the code, we consider the modeling of light curves, flux centroid proper motion, and image size. 

To start with, we consider an observer who performs the actual observation, referred to as the lab observer. It is assumed that the lab observer's LOS lies within the x-z plane of the jet's coordinate system, and is offset from the z-axis by a polar angle $\theta_{\rm v}$. From the lab observer's perspective, the jet's image is projected onto a plane perpendicular to the LOS. To model the size of this image, we establish a two-dimensional Cartesian coordinate system, $\tilde{x}-\tilde{y}$, within this plane, referred to as the LOS coordinate system. This LOS coordinate system is defined so that the jet's motion is oriented along the $\tilde{x}$ direction, and the $\tilde{y}$ axis aligns with the y-axis of the jet's coordinate system.

\subsection{Relativistic effects} \label{subsec:equations}
When modeling the radiation from a relativistic expanding shell, two major relativistic effects must be considered: the Doppler beaming and the time difference in arrival between different latitudes. Due to the spreading and nonzero tangent velocity, the calculation of these effects involves some complexities, which we will describe below.

We begin by addressing the arrival time difference effect. When the jet's expansion speed is ultrarelativistic, the emission of fluid elements from various latitudes that appear to arrive simultaneously at the lab observer is not actually emitted at the same time in the burster frame. The geometric shape that characterizes the surface where emissions reach the lab observer at the same time is commonly referred to as the ``equal arrival time surface," and is defined by the following equation:
\begin{equation}\label{eq:eats}
    t - R\mu_{\rm r}/c = T_{\rm obs}/(1 + z).
\end{equation}
In this equation, $\mu_{\rm r}$ is the cosine angle between the radial direction and the LOS, $T_{\rm obs}$ is the observed time of radiation since the initial burst, and $1+z$ is the time dilation factor due to cosmological redshift. The cosine angle can be calculated by the inner product of radial norm $\hat{r}$ and LOS norm $\hat{n}$:
\begin{align}
    \mu_{\rm r} & = \hat{r}\cdot\hat{n} \nonumber \\
    & = \cos\theta\cos\theta_{\rm v} + \sin\theta\cos\phi\sin\theta_{\rm v}
\end{align}
In the case of an isotropic shell expanding at a constant speed, the equal arrival time surface is an ellipsoid. However, for a structured jet, the shape becomes nontrivial and must be solved numerically. One must convert from $T_{\rm obs}$ to $t$ by solving this equation to derive the actual emission time in the burster frame. 

To efficiently solve the equation, we first use a binary search algorithm to identify the range of the solution within the discrete numerical data. The data is further linear interpolated in this range to approximate $R(t,\theta)$. This routine has significantly improved the performance because a root solver is no longer needed in this step.

When considering lateral expansion, potential peculiar solutions could arise if the normal vector of a relativistic blast wave significantly deviates from the radial direction. This situation might lead to the coordinate speed $\partial R/\partial t$ exceeding the speed of light. While this does not reflect a problem for hydrodynamics because it does not represent the actual speed of a fluid element, it does introduce numerical challenges in solving the equal arrival time surface because a fluid element from one direction can ``catch up with" a light emitted from another direction, which leads to multiple solutions of equal arrival time surface at a given pair of $T_{\rm obs}$ and $\theta$.

In this work, we do not account for this effect and assume that eq. \ref{eq:eats} has only one numerical solution given a pair of $T_{\rm obs}$ and $\theta$, because in practice we have not observed multiple solutions in any of the numerical setups that we have tested thus far. This might happen because spreading happens only in the mildly relativistic regime. However, we leave this caveat to acknowledge that it could potentially present a numerical challenge in certain scenarios.

Another relativistic effect is the Doppler beaming, which is characterized by the Doppler factor
\begin{equation}
    \delta = \frac{1}{\gamma(1-\beta\mu_{\beta})}.
\end{equation}
Here, $\mu_{\beta}$ represents the cosine angle between the velocity of the fluid element and the direction of the LOS. Due to the nonzero polar velocity, we cannot substitute $\mu_{\beta}$ with $\mu_{\rm r}$ as is done in many semianalytical models. The exact value of $\mu_{\beta}$ can be calculated by taking the inner product of velocity norm $\hat{\beta}$ and LOS norm $\hat{n}$ like above. This calculation yields
\begin{align}
    \mu_{\beta} & = \hat{\beta}\cdot\hat{n} \nonumber \\
                & = (\frac{\beta_{\rm r}}{\beta}\sin\theta\cos\phi + \frac{\beta_{\theta}}{\beta}\cos\theta\cos\phi)\sin\theta_{\rm v} \nonumber \\
                & + (\frac{\beta_{\rm r}}{\beta}\cos\theta - \frac{\beta_{\theta}}{\beta}\sin\theta)\cos\theta_{\rm v}.
\end{align}
Numerical tests have revealed minor differences between using the exact expression and approximating it with $\mu_{\rm r}$. This can be attributed to the fact that spreading primarily occurs when the jet reaches mildly relativistic speeds, and the beaming effect is relatively unimportant during this phase.

\subsection{Light Curves} \label{subsec:lc}
We now proceed to model the afterglow light curves from a relativistic jet, which is one of the most important observational outcomes of GRB afterglow. For simplicity, here we consider a minimal assumption of synchrotron radiation following \citealt{1998ApJ...497L..17S}. We briefly summarize it as follows.

We assume that the electrons accelerated by the blast wave follow a nonthermal power-law distribution with a power index of $-p$. The minimal Lorentz factor $\gamma_{\rm m}$ of electrons can be determined by assuming that a fraction $\epsilon_{\rm e}$ of the internal energy is converted into electron kinetic energy. The result is expressed as:
\begin{equation}\label{eq:gamma_m}
    \gamma_{\rm m} = \frac{p - 2}{p - 1}\frac{\epsilon_{\rm e}m_{\rm p}}{m_{\rm e}}(\gamma - 1)
\end{equation}

The relativistic electrons are assumed to move in a randomly oriented magnetic field. The average magnetic field can be calculated by assuming that a fraction $\epsilon_{\rm B}$ of the internal energy is converted into magnetic energy:
\begin{equation}
    B' = \sqrt{8\pi \bar{e}'_{\rm sw}\epsilon_{\rm B}}
\end{equation}
where $n_0=\rho_0/m_{\rm p}$ is the number density of external medium. 

The synchrotron radiation spectrum can be approximated by a multiband broken power law, which depends on several characteristic break frequencies. In this work, we consider two such breaks: the frequencies emitted by electrons with the minimum Lorentz factor $\gamma_{\rm m}$ and the cooling frequency associated with the Lorentz factor $\gamma_{\rm c}$. For simplicity, we do not consider synchrotron self-absorption. The cooling Lorentz factor is calculated via global cooling approximation
\begin{equation}
    \gamma_{\rm c}=\frac{6\pi m_{\rm e}\gamma c}{\sigma_{\rm T}B'^2t}
\end{equation}
The break frequencies can be calculated by
\begin{equation}
    \nu'_{\rm i} = \frac{3eB'\gamma_{\rm i}^2}{4\pi m_{\rm e}c}
\end{equation}
where ${ i}=\{ {\rm m}, {\rm c}\}$ and $e$ is the electron charge. The synchrotron emissivity in the comoving frame is now expressed by
\begin{align}
	& \epsilon'_{\nu'} = \epsilon'_P \nonumber \\
    & \times
	\begin{cases}
		(\nu' / \nu'_m)^{1/3} & \nu' < \nu'_m < \nu'_c \\
		(\nu' / \nu'_m)^{-(p-1)/2} & \nu'_m < \nu' < \nu'_c \\
		(\nu'_c / \nu'_m)^{-(p-1)/2} (\nu' / \nu'_c)^{-p/2} & \nu'_m < \nu'_c < \nu' \\
		(\nu' / \nu'_m)^{1/3} & \nu' < \nu'_c < \nu'_m \\
		(\nu' / \nu'_c)^{-1/2} & \nu'_c < \nu' < \nu'_m \\
		(\nu'_m / \nu'_c)^{-1/2} (\nu' / \nu'_m)^{-p/2} & \nu'_c < \nu'_m < \nu'
	\end{cases}
\end{align}
The peak emissivity can be calculated by
\begin{equation}
    \epsilon'_{\rm P} =\frac{\sqrt{3}e^3B'n'}{m_{\rm e}c^2}
\end{equation}
where $n'=\rho'_{\rm sw}/m_{\rm p}$ is the number density of the blast.

The total luminosity can be determined by integrating the emissivity over the radiating volume. In our model, because the radiating region is approximated as an infinitely thin surface, we need to estimate a typical shell width to perform the volume integral. This estimation can be accomplished by relating the mass per solid angle ($M_{\rm sw}$) with the mass per volume ($\rho'_{\rm sw}$) and realizing that $dV=r^2drd\Omega$. As such, the typical shell width can be approximated as
\begin{equation}
    \Delta R \approx \frac{M_{\rm sw}}{R^2\rho'_{\rm sw}\gamma}
\end{equation}
where we recover $\rho'_{\rm sw}$ from the delta function approximation to the value determined by the shock jump condition.

Given the estimation of typical shell width, the total luminosity can be estimated by the integral
\begin{equation}\label{eq:luminosity}
    L_{\nu} = \int  4\pi I_{\nu} R^2 d\Omega,
\end{equation}
where $I_{\nu}=I'_{\nu'}\delta^3$ is the specific intensity and $\nu = \nu'\delta/(1+z)$ is the observed frequency. The specific intensity can be estimated by
\begin{equation}
    I'_{\nu'} = \int\frac{\epsilon'_{\nu'}}{4\pi}dr' \approx \frac{\epsilon'_{\nu'}}{4\pi}\Delta R',
\end{equation}
where $\Delta R'=\Delta R\gamma$ is the shell width in the comoving frame. The integral is performed in the jet spherical coordinate at lab observer's time $T_{\rm obs}$. The time of radiation $t$ for each fluid element is solved by the equal arrival time surface discussed above.

The integral is carried out using a two-dimensional adaptive integration algorithm. Due to the relativistic beaming, the integrand exhibits a very narrow peak. To efficiently and accurately perform the integration, we first employ a minimization algorithm to identify the peak. After that, we rotate the spherical coordinates to align the peak with the north pole. This approach takes advantage of the fact that polar lines converge at the poles, which naturally results in an oversampling of integration points around the peak, leading to improved integration efficiency. These implementations offer a significant enhancement in performance compared to a direct integration in the jet coordinate system.

Given the luminosity, we can calculate the flux density by
\begin{equation}
    F_{\nu} = \frac{L_{\nu}(1+z)}{4\pi D_{\rm L}^2}
\end{equation}
where $D_{\rm L}$ is the luminosity distance. 

The above minimal synchrotron radiation model is set as default in \texttt{jetsimpy} and is sufficient for afterglow modeling in most cases. Nevertheless, the flexibility of the code allows for convenient customization with more sophisticated modeling.

\subsection{Centroid proper motion and image size} \label{subsec:superluminal}
Another important observable of GRB afterglow is the flux centroid proper motion, which provides a direct constraint of the Lorentz factor. The proper motion of the jet image in the LOS plane can be defined by the offset of a ``flux centroid" from the burst origin. Since the image of the jet has a diffusive scale, the overall offset can be defined by a weighted average over the intensity of the image (e.g., \citealt{2018ApJ...865L...2Z, 2022MNRAS.509..395F, 2022Natur.610..273M, 2023MNRAS.524.5514N}). Therefore, the flux centroid position in the LOS plane can be calculated by
\begin{equation}
    \tilde{x}_{\rm c} = \frac{\int R\lambda_{\rm r} \frac{dL_{\nu}}{d\Omega}d\Omega}{\int \frac{dL_{\nu}}{d\Omega}d\Omega}
\end{equation}
where $dL/d\Omega$ is defined in eq.\ref{eq:luminosity}, and $\lambda_{\rm r}$ is the sine angle between the radial direction and the LOS. It can be calculated by
\begin{equation}
    \lambda_{\rm r} = - \sin\theta\cos\phi\cos\theta_{\rm v} + \cos\theta\sin\theta_{\rm v}
\end{equation}

The angular offset from the burst origin can be calculated by
\begin{equation}
    \tilde{\theta}_{\rm c} = \frac{\tilde{x}_{\rm c}}{D_{\rm A}}
\end{equation}
where $D_{\rm A} = D_{\rm L}/(1+z)^2$ is the angular diameter distance. 

We can also calculate the image size of the jet similarly. Following the definitions adopted in previous works (e.g., \citealt{2023arXiv231002328R}), the image size is calculated by taking the moments of the intensity distribution with respect to the flux centroid. The Gaussian-equivalent image sizes along the $\tilde{x}$ and $\tilde{y}$ directions are defined as follows:
\begin{align}
    \tilde{\sigma}^2_{\rm x} & = \frac{\int(R\lambda_{\rm r} - \tilde{x}_{\rm c})^2\frac{dL_{\nu}}{d\Omega}d\Omega}{\int\frac{dL_{\nu}}{d\Omega}d\Omega} \\
    \tilde{\sigma}^2_{\rm y} & = \frac{\int (R\sin\theta\sin\phi-\tilde{y}_{\rm c})^2 \frac{dL_{\nu}}{d\Omega}d\Omega}{\int \frac{dL_{\nu}}{d\Omega}d\Omega}
\end{align}
Given the axial symmetry of the jet, we have $\tilde{y}_{\rm c}=0$. The angular sizes are then calculated as:
\begin{align}
    \tilde{\theta}_{\rm x} & = \frac{\tilde{\sigma}_{\rm x}}{D_{\rm A}} \\
    \tilde{\theta}_{\rm y} & = \frac{\tilde{\sigma}_{\rm y}}{D_{\rm A}}
\end{align}

\section{Tests of Afterglow Modeling} \label{sec:test}
In this section, we test our afterglow modeling described in the previous section by comparing them to existing tools. For the light-curve calculation, we compare our code with the widely-used tools \texttt{afterglowpy} \citep{2020ApJ...896..166R} and \texttt{BoxFit} \citep{2012ApJ...749...44V}. These tools represent the state-of-the-art modeling of GRB afterglow with semianalytical methods or full numerical hydrodynamics. For the centroid proper motion and image size, we directly compare them to the simulated imaging of the GRB jet.

\subsection{Comparison to \texttt{afterglowpy}} \label{subsec:afterglowpy}

The package \texttt{afterglowpy} \citep{2020ApJ...896..166R} is a widely-used GRB afterglow modeling tool with a semianalytical spreading prescription. In this model, a structured jet is approximated by the combination of various concentric top-hat jets moving in an ISM environment (see also \citealt{2017MNRAS.472.4953L, 2018MNRAS.481.2581L,2022MNRAS.509..395F,2023MNRAS.520.2727N} with a similar approach). Each jet segment laterally expands at local sound speed. The spreading is initially frozen until the jets become mildly relativistic. This semianalytical approach provides a reasonable approximation of the spreading phase at a much lower computational cost compared to hydrodynamic simulations. However, because the interactions between the fluid elements are neglected, it can not account for all nonlinear hydrodynamic effects or achieve the same level of accuracy as a full simulation.

Our model shares the same trans-relativistic equation of state and thin shell approximation as in \texttt{afterglowpy}, as well as a very similar radiation calculation. The major differences are our more flexible initial angular jet profile, the inclusion of the coasting phase, a different spreading prescription, the calibration with self-similar solutions, and the support for the wind environment. In fact, if all these effects are disabled, the numerical results of the blast wave evolution in the two models should agree {\it exactly} the same. 

To compare the light curves for a structured jet, we assume the isotropic equivalent energy and initial Lorentz factor follow a Gaussian distribution with a typical core width $\theta_{\rm c}$:
\begin{align}
    E(\theta) & = E_0 \exp{\left[-\frac{1}{2}\left(\frac{\theta}{\theta_{\rm c}}\right)^2\right]} \\
    \gamma(\theta) & = (\gamma_0 - 1)\exp{\left[-\frac{1}{2}\left(\frac{\theta}{\theta_{\rm c}}\right)^2\right]} + 1.
\end{align}

\begin{figure}[ht!]
\centering
    \includegraphics[width=\columnwidth]{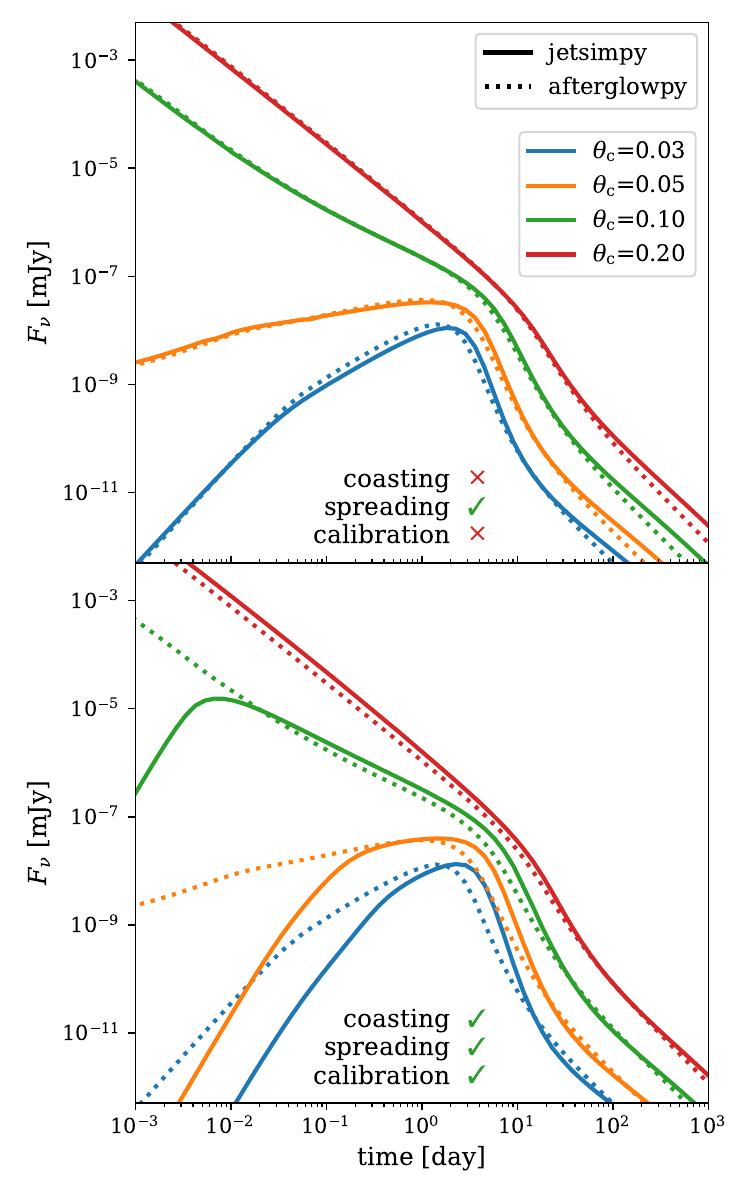}
\caption{Comparison of light curves to the publicly available code \texttt{afterglowpy}. The jet structures are assumed to be Gaussian profiles with core width $\theta_{\rm c}$ labeled in the figure. The rest of the parameters for the light curves are $n_0=1$ cm$^{-3}$, $E_0=10^{51}$ erg, $\gamma_0=1000$, $\theta_{\rm v}=0.3$, $\epsilon_{\rm e}=0.1$, $\epsilon_{\rm B}=0.01$, $p=2.5$, $\nu=10^{18}$ Hz, $d_{\rm L}=474.33$ Mpc, and $z=0.1$. Top panel: light curves with spreading effect and calibration disabled ($\gamma_0=\infty$ and $s=1$). Bottom panel: light curves with these effects enabled.
\label{fig:compare_afterglowpy_gaussian}}
\end{figure}

The comparisons of light curves are presented in Fig. \ref{fig:compare_afterglowpy_gaussian}, with the parameters provided in the caption. In this comparison, we keep the viewing angle fixed at $\theta_{\rm v}=0.3$ and vary the half-opening angle $\theta_{\rm c}$. To account for radiation from all directions, we set the truncation angle in \texttt{afterglowpy} to $\theta_{\rm w}=\pi/2$. 

In the upper panel, we enable only the spreading effect, while disabling the coasting phase and calibration (i.e., $\gamma_0=\infty$ and $s=1$). This comparison is intended to isolate and reveal the differences caused solely by the modeling of the spreading effect. We observe overall good agreement in all cases, suggesting that nonlinear hydrodynamic effects are not significant for structured jets. However, we do notice minor discrepancies, especially for narrow jets. Specifically, our model predicts a slightly faster spreading effect than \texttt{afterglowpy} for very narrow jets (e.g., the blue line). Such rapid spreading is expected in these cases because the shock front in narrow jets is more causally connected, leading to an exponential expansion similar to the prescription in \citealt{1999ApJ...525..737R}. This fast-spreading effect for extremely narrow jets has also been observed in many numerical hydrodynamic studies (e.g., \citealt{2012MNRAS.421..570G,2023arXiv231109297G,2023MNRAS.520.2727N}). The minor discrepancy observed at later times is likely due to our different spreading prescription because this discrepancy disappears when spreading is disabled.

In the lower panel, we maintain the same numerical setup, with the addition of enabling both the coasting effect and calibration. The initial Lorentz factor of the core is assumed to be 1000. This finite initial Lorentz factor leads to an initial rising phase following a slope of $F_{\nu}\propto t^3$, resulting in a significant difference for the early time. The effect of calibration is approximately a temporal offset because the coefficient can be equivalently understood as a scaled external density. We also find the calibration leads to a slightly faster decay in the spreading phase. This effect is due to the transition of the calibration coefficients between the self-similar solutions in each limit. In the relativistic limit, this coefficient is less than 1 (see Fig. \ref{fig:calib}), resulting in a predicted velocity that is faster than uncalibrated predictions. Conversely, in the Newtonian limit, the coefficient is greater than 1, leading to a lower velocity prediction. As a result, the combined prediction of velocity decreases more rapidly in the transitional regime than in either of the limits, which occurs exactly during the spreading phase.

\subsection{Comparison to \texttt{BoxFit}} \label{subsec:boxfit}
We further compare to the light curves that have been post-processed using full numerical hydrodynamics; specifically, the widely-used tool \texttt{BoxFit} \citep{2012ApJ...749...44V}. This code employs top-hat angular structures and a Blandford-McKee radial profile for its initial setup. The simulation data are then interpolated for post-processing the light curve and spectrum modeling. \texttt{BoxFit} calculates the observed flux density through a ray-tracing radiative transfer module. Because the code is based on numerical simulations, the synthetic light curves are considered highly reliable. However, the code only works for a top-hat jet structure. Moreover, \texttt{BoxFit} supports a wind external medium, making it a unique tool for comparisons in such scenarios.

\begin{figure}[ht!]
\centering
    \includegraphics[width=\columnwidth]{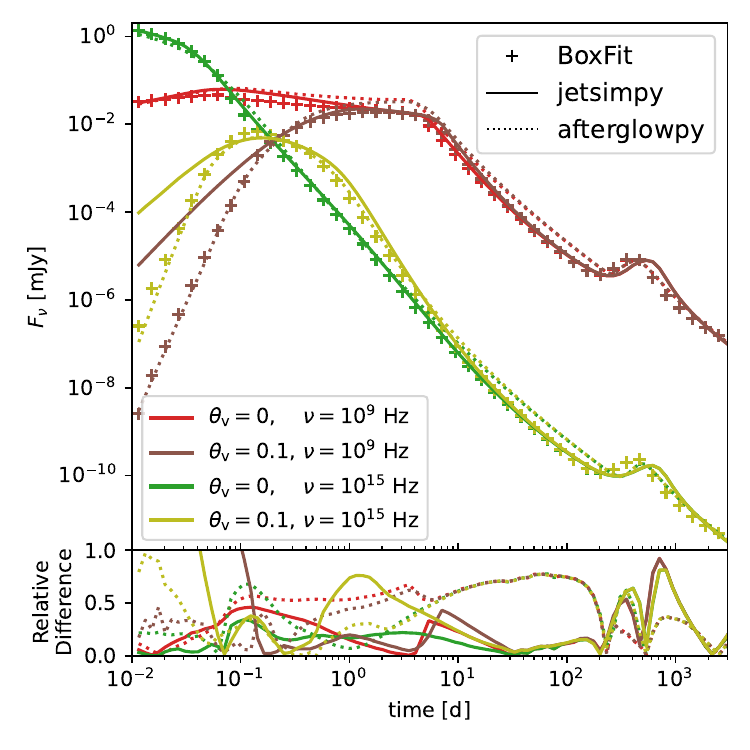}
\caption{Comparison of light curves to the publicly available codes \texttt{BoxFit} and \texttt{afterglowpy}. We assume a top-hat jet with an isotropic equivalent energy of $10^{52}$ erg, a half-opening angle of 0.045 radians (3 degrees), and a symmetric counter-jet. The initial Lorentz factor is infinite and the external medium is ISM-like. The rest of the parameters are: $n_0=1$ cm$^{-3}$, $E_{\rm iso}=10^{52}$ erg, $\epsilon_{\rm e}=0.1$, $\epsilon_{\rm B}=0.01$, $p=2.5$, $d_{\rm L}=10^{28}$ cm, and $z=0$. The viewing angle $\theta_{\rm v}$ and the frequency $\nu$ are labeled in the figure. Top panel: Radio and optical light curves. Bottom panel: Relative differences as compared to \texttt{BoxFit}.
\label{fig:compare_boxfit_ism}}
\end{figure}

In Figure \ref{fig:compare_boxfit_ism}, we present the comparisons of radio and optical light curves for both on-axis and off-axis observations in an ISM environment. Considering that nonlinear hydrodynamic effects are more significant in narrower jets, we performed a test to evaluate the reliability of our code under these extreme conditions. For this purpose, the comparison was carried out using the minimum jet half-opening angle allowed by \texttt{BoxFit}, which is 0.045 radians. To align our numerical setup with \texttt{BoxFit}, we disable the coasting phase while enabling the spreading effect and calibration. Additionally, we add a counter-jet to assess the robustness of our model under Newtonian limits. As a reference, the light curves produced by \texttt{afterglowpy} are also presented. Details of the jet parameters and light curves are provided in the figure caption.

Not surprisingly, our results show good agreement with \texttt{BoxFit}, typically within a factor of 50\%. In particular, the agreement in the spreading phase is slightly better than \texttt{afterglowpy}. The major discrepancies observed in this comparison can be explained as follows. The significant difference at early times for off-axis cases (brown and olive lines) is attributed to the bow shock (see \S \ref{sec:jet_evolution}) generated outside the jet's core. This effect is particularly significant in a top-hat jet due to its very sharp edge. In \texttt{BoxFit} simulations, these periods are earlier than the startup time, and the calculations are based on extrapolations assuming a Blandford-McKee radial profile and top-hat angular profile, where bow shocks are not present. The rising times of counter-jets also show a temporal offset. This suggests that the assumption of an infinitely thin shell may not hold in the Newtonian limit, as we have anticipated. However, the temporal difference remains within a factor of 30\%. Additionally, the early spreading phase for off-axis observation (the olive line) shows a minor deviation, likely due to the more extended tail discussed in \S \ref{sec:compare_hydro}, but the difference remains within an order of unity. This comparison demonstrates that our model can achieve an overall accuracy comparable to that of full numerical hydrodynamics, even when nonlinear effects are significant.

\begin{figure}[ht!]
\centering
    \includegraphics[width=\columnwidth]{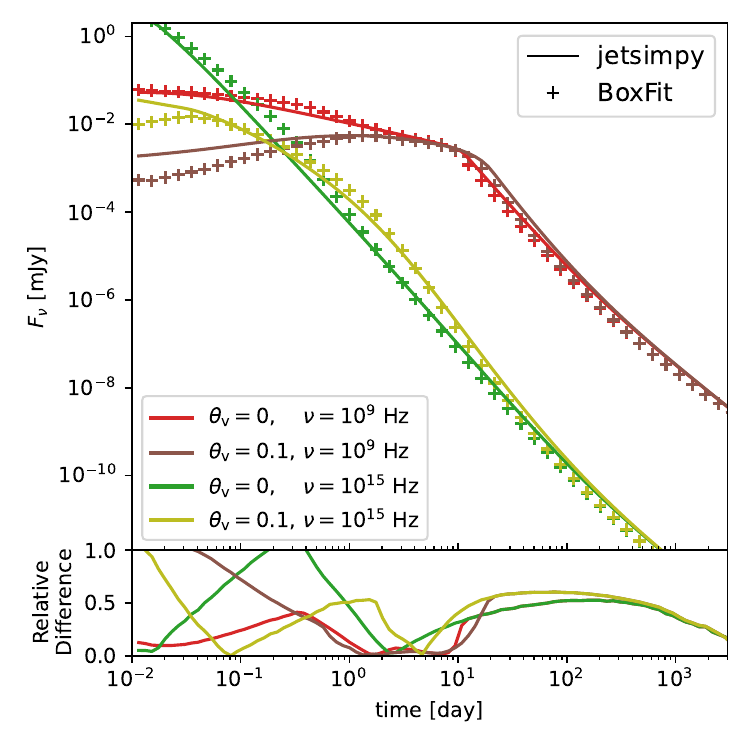}
\caption{Same as Figure \ref{fig:compare_boxfit_ism}, but for a wind-like environmental medium profile: $n_0=1(r/10^{17}{\rm cm})^{-2}$ cm$^{-3}$. The counter-jet is absent in this comparison.
\label{fig:compare_boxfit_wind}}
\end{figure}

In Figure \ref{fig:compare_boxfit_wind}, we present the comparison of our model with \texttt{BoxFit} in a wind environment. For simplicity, we do not include counter-jets in this comparison because the \texttt{BoxFit} simulations in these cases are performed in a Lorentz boosted frame, which does not resolve counter-jets effectively. This comparison also demonstrates overall agreement, typically within a factor of unity, but is somewhat less precise than in ISM environments, as we have expected. The deviations at the early time are still due to the bow shocks that emerged outside the jet core. It is worth noting that in the wind environment, this level of agreement is only attainable when calibration is enabled. Without calibration, the results will exhibit a significant temporal offset. This suggests that the calibration coefficient has a bigger impact on solutions in wind environments as compared to ISM environments.

\subsection{Tests of Flux Centroid Proper Motion and Image Size Modeling} \label{subsec:verify_centroid_size}
To validate our calculation of the centroid proper motion and image size, we directly compare it to simulated images generated from the numerical data. The procedure for calculating afterglow imaging is summarized as follows. For each pixel of the image we consider a ray parallel to the LOS but offset from the origin by a corresponding distance. This ray is then intersected with the equal arrival time surface. At each intersection point, we approximate the observed intensity using $I_{\nu}=\epsilon'_{\nu'}\Delta R'\delta^3/4\pi$. The total intensity for a given pixel is calculated as the sum of intensities from all intersecting points. The image reflects the observed intensity distribution within the LOS plane.

\begin{figure}[ht!]
\centering
    \includegraphics[width=\columnwidth]{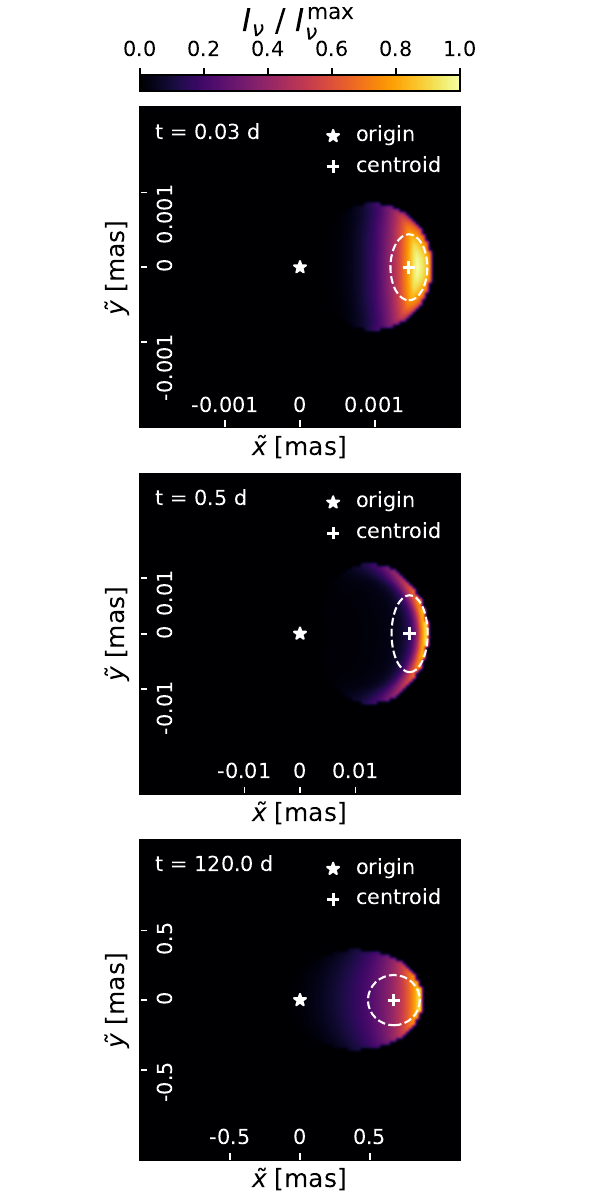}
\caption{An image of a jet with a Gaussian profile traveling in an ISM environment. From top to bottom are the images at the coasting phase, deceleration phase, and spreading phase. The intensity distributions are normalized by maximum intensity. The markers ``$\star$" and ``$+$" represent the burst origin and the flux centroid, respectively. The white contour is the $1-\sigma$ size assuming Gaussian-equivalent intensity distribution. The parameters for the calculation are: $n_0=1$ cm$^{-3}$, $E_0=10^{52}$ erg, $\gamma_0=1000$, $\theta_{\rm c}=0.1$ rad, $\theta_{\rm v}=0.3$ rad, $\epsilon_{\rm e}=10^{-2}$, $\epsilon_{\rm B}=10^{-3}$, $p=2.2$, $\nu=3\times10^{9}$ Hz, $d_{\rm L}=44$ Mpc, and $z=0.01$.
\label{fig:image}}
\end{figure}

Our results are presented in Fig. \ref{fig:image} with parameters detailed in the caption. The images, arranged from top to bottom, correspond to three typical stages of the jet: the coasting phase, the deceleration phase, and the spreading phase. The flux centroid locations are marked with ``+" signs, and the image sizes are presented by the $1-\sigma$ contours of the Gaussian-equivalent intensity distribution. Our modeling of the flux centroid and image size closely aligns with the simulated images, thereby validating the accuracy of our method.

\section{Application to GRB 170817A afterglow}\label{sec:observation}

In this section, our model is applied to real-world data of the GRB 170817A afterglow following the gravitational wave event GW170817 \citep{2017PhRvL.119p1101A, 2017ApJ...848L..12A, 2017ApJ...848L..13A}. This event represents the first multimessenger observation of neutron star mergers. It is also the first GRB afterglow whose relativistic jet is evidently observed from an off-axis direction (see \citealt{2021ARA&A..59..155M} for a review). The multiwave band observation spanning a significant duration, together with the well-measured apparent superluminal motion \citep{2018Natur.561..355M,2019Sci...363..968G,2022Natur.610..273M}, makes it an ideal object to apply our model.

To better fit the data of this event, we have made a slight adjustment to the synchrotron radiation model described in \S \ref{subsec:lc}, incorporating the correction of the ``deep Newtonian phase" at late stages \citep{2003MNRAS.341..263H,2006ApJ...638..391G,2013ApJ...778..107S}. This adjustment is necessary when the minimal Lorentz factor of electrons, $\gamma_{\rm m}$, as calculated by Eq. \ref{eq:gamma_m}, drops below 1, which happens when the velocity of the blast wave has sufficiently decreased. In this scenario, electrons follow a power-law distribution in momentum space, rather than in energy space. Consequently, only a subset of electrons are relativistic and contribute to the emission in the GHz wave band. This adjustment might be necessary given that a recent study \citep{2023arXiv231002328R} has shown it to yield improved fits for late-time data. Here, we follow the approach described in \citealt{2013ApJ...778..107S}. Specifically, if $\gamma_{\rm m}$ falls below 1, we fix it at 1 and reduce the number of radiating electrons by a fraction of $f$:
\begin{equation}
    f = \frac{p - 2}{p - 1} \frac{\epsilon_{\rm e} m_{\rm p}} {\gamma_{\rm m} m_{\rm e}}(\gamma-1).
\end{equation}
This correction leads to a late-time flattening of the afterglow light curve.

We emphasize that the objective of this section is not to conduct a comprehensive study of GRB 170817A nor GW170817, but rather to demonstrate the application of our code using real-world data. As such, we do not explore the effects of varying jet structures or the potential influence of additional phenomena, such as kilonova afterglows \citep{2011Natur.478...82N, 2013MNRAS.430.2121P, 2017ApJ...848L..21A,2018ApJ...867...95H,2019MNRAS.487.3914K}, that may emerge at later stages. Furthermore, gravitational wave data is not incorporated into our analysis, given its irrelevance to afterglow modeling. A comprehensive study of multimessenger data of GW170817 using our newly developed code is left for future research.

\subsection{data fitting}

To fit the data with our model, we perform a Bayesian parameter estimation. For this analysis, multiwave band light-curve data up to 1000 days are taken from \citet{2021ApJ...922..154M}, where the data have been re-analyzed to minimize discrepancies arising from different data processing techniques. The original data references are detailed within their work. Beyond 1000 days, we incorporate radio wave observations from the Very Large Array \citep{2021ApJ...914L..20B} and X-ray measurements from Chandra \citep{2021ApJ...914L..20B,2022ApJ...938...12B,2022GCN.32065....1O}. Additionally, we take into account the flux centroid positions measured by the Very Large Baseline Interferometry (VLBI; \citealt{2018Natur.561..355M,2019Sci...363..968G}) and the Hubble Space Telescope (HST; \citealt{2022Natur.610..273M}), with the HST measurement presumed to mark the origin of the burst. Since there is a correlated systematic error between the VLBI and HST measurements due to different coordinate systems, for simplicity this error is integrated into the HST data point. The total uncertainty for each offset measurement is calculated as the root mean square of the statistical and systematic errors. Since gravitational wave data is excluded from our analysis, we set the luminosity distance at $D_{\rm L}$=43.9 Mpc and the redshift at $z=0.0098$, assuming cosmological parameters adopted from Planck \citep{2020A&A...641A...6P}. To simplify the analysis, data points representing upper limits are omitted from all mentioned datasets.

In this study, we adopt a Gaussian profile for the angular distribution of the jet, as described in the previous section. The absence of an obvious coasting phase leads us to presume an infinitely large initial Lorentz factor. The lack of evident late-time re-brightening at the current stage motivates us to exclude the counter-jet component (e.g. \citealt{2024arXiv240117978L}). To fit the apparent superluminal motion, we assume that the trajectory of the flux centroid within the LOS plane originates from the coordinates $(RA_0, Dec_0)$ and with an orientation angle of $\alpha$. Similar to previous studies \citep{2018Natur.561..355M,2019Sci...363..968G,2022Natur.610..273M,2024MNRAS.528.2600G,2023arXiv231002328R}, we place the coordinate origin at the centroid's position at 75 days. Under these assumptions, we identify the following set of free parameters for our model fitting: $\vec{\vartheta}$=\{$\log_{10}n_0$, $\log_{10}E_0$, $\theta_{\rm c}$, $\theta_{\rm v}$, $\log_{10}\epsilon_{\rm e}$, $\log_{10}\epsilon_{\rm B}$, $p$, $RA_0$, $Dec_0$, $\alpha$\}.

According to the Bayes' theorem, the posterior probability distribution of the fitting parameters can be expressed as:
\begin{equation}
p(\vec{\vartheta} | \vec{d}_{\rm F}, \vec{d}_{\rm c}) \propto \mathcal{L}(\vec{d}_{\rm F}|\vec{\vartheta})\mathcal{L}(\vec{d}_{\rm c}|\vec{\vartheta})\pi(\vec{\vartheta}),
\end{equation}
where $\vec{d}_{\rm F}$ and $\vec{d}_{\rm c}$ represent the datasets for flux density and flux centroid offset, respectively. The likelihoods $\mathcal{L}$ for both datasets are modeled as $\exp(-\chi^2/2)$, with $\chi^2$ being the reduced chi-squared value calculated from the discrepancies between observed data and the model predictions. It has been observed by \citealt{2023arXiv231002328R} that the late-time X-ray observations for this event, due to their low photon counts, would be better modeled using a Poisson likelihood. This adjustment has not been implemented in our current analysis but is subject to future exploration. The prior distribution $\pi(\vec{\vartheta})$ for the parameters is detailed in Table \ref{tab:prior}.

\begin{table}
	\centering
	\caption{Prior distributions of fitting parameters.}
	\label{tab:prior}
	\begin{tabular}{ccc} 
		\hline
		Parameters & Distributions & Bounds\\
		\hline
		$\log_{10}(n_0/{\rm cm}^{-3})$ & Uniform & [-5, 0]\\
            $\log_{10}(E_0/{\rm erg})$ & Uniform & [49, 57] \\
            $\theta_{\rm c}$ [rad] & Uniform & [0.01, $\pi/2$] \\
            $\theta_{\rm v}$ [rad] & Sine & [0, $\pi$/2] \\
            $\log_{10}\epsilon_{\rm e}$ & Uniform & [-6, 0] \\
            $\log_{10}\epsilon_{\rm B}$ & Uniform & [-6, 0] \\
            $p$ & Uniform & [2, 2.5] \\
            $RA_0$ & Uniform & [-10, 10] \\
            $Dec_0$ & Uniform & [-10, 10] \\
            $\alpha$ & Uniform & [-$\pi$, $\pi$] \\
            \hline
	\end{tabular}
\end{table}

In this analysis, we conduct two separate MCMC studies: one that incorporates all aforementioned datasets, and another that excludes the centroid offset data. To sample the posterior probability distribution, we utilize the \texttt{Bilby} package, a public tool for Bayesian data analysis \citep{2019ApJS..241...27A}. Within \texttt{Bilby}, we employ the \texttt{pymultinest} sampler \citep{2014A&A...564A.125B}, which is based on a Nested Sampling algorithm. Nested Sampling offers an improved computational efficiency for high-dimensional problems. 

\begin{figure}[ht!]
\centering
    \includegraphics[width=\columnwidth]{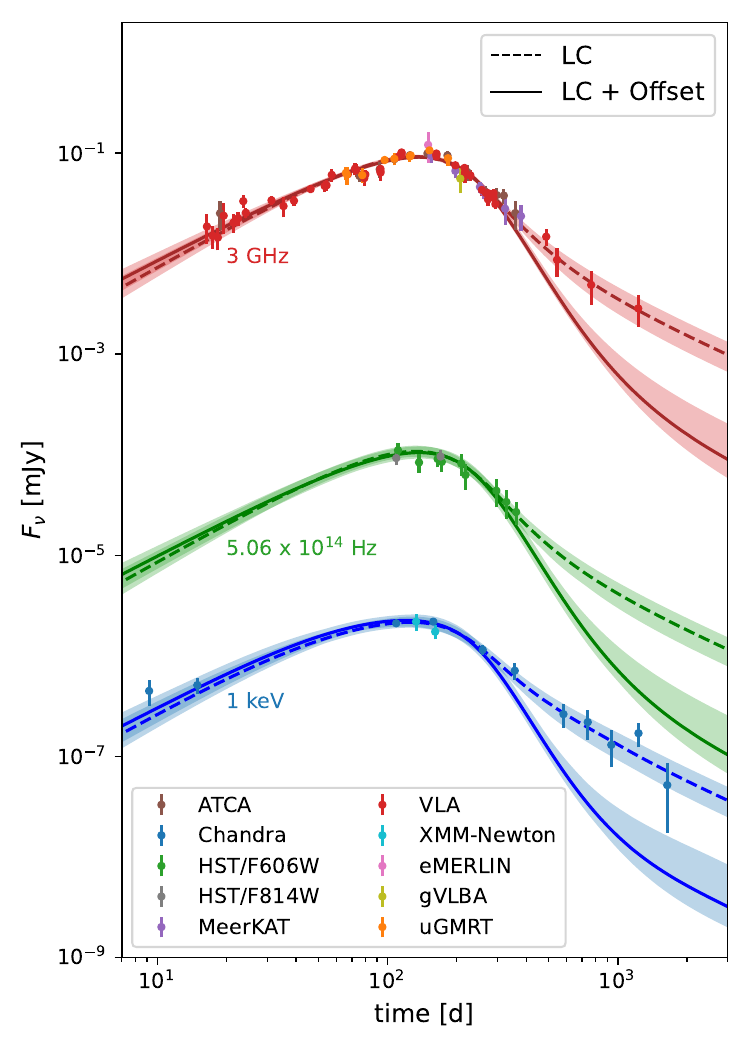}
\caption{Best fitting light curves for multiwave band observation of GRB 170817A afterglow with flux centroid data included or not. The data points are normalized to 3 GHz, 5.06$\times 10^{14}$ Hz, and 1keV, respectively. The shadow regions represent a 68\% confidence interval of the flux density.
\label{fig:lc_fit}}
\end{figure}

\begin{figure}[ht!]
\centering
    \includegraphics[width=\columnwidth]{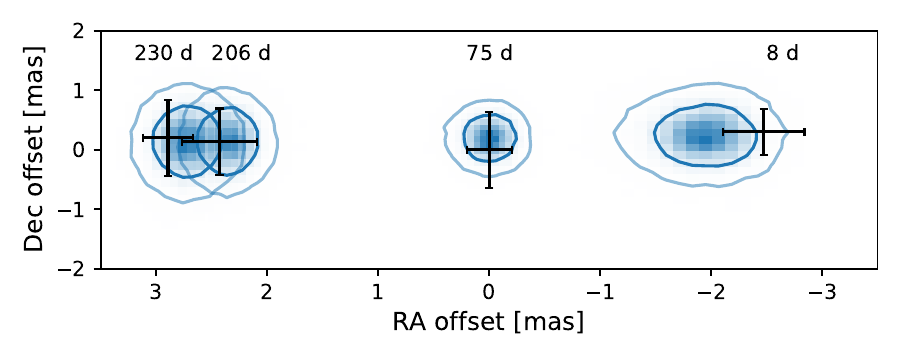}
\caption{The posterior distribution of flux centroid position with respect to the measurement data. The contour lines represent 68\% and 95\% confidence regions. The measurement at 8 days is assumed to be the burst origin.
\label{fig:offset_fit}}
\end{figure}

In Fig. \ref{fig:lc_fit}, we present the light curves generated using the sample with the highest posterior probability, together with the 68\% confidence intervals for the flux density. The fitting results for the centroid offset are shown in Figure \ref{fig:offset_fit}. We summarize the parameter estimation results in Table \ref{tab:posterior}. The full corner plots of posterior probability distributions are shown in Figure \ref{fig:corner} and Figure \ref{fig:corner_offset} in Appendix \ref{appendix:posterior}.

\begin{table}
    \centering
    \caption{Parameter estimation results for the two MCMC fittings.}
    \label{tab:posterior}
    \begin{tabular}{ccc} 
    \hline
	Parameters & LC & LC + Offset\\
	\hline
	$\log_{10}(n_0/{\rm cm}^{-3})$ & $-0.65^{+0.32}_{-0.36}$ & $-1.33^{+0.91}_{-1.03}$\\
        $\log_{10}(E_0/{\rm erg})$ & $51.86^{+0.41}_{-0.48}$ & $54.53^{+0.92}_{-0.95}$ \\
        $\theta_{\rm c}$ [deg] & $7.55^{+0.62}_{-0.59}$ & $2.84^{+0.29}_{-0.23}$ \\
        $\theta_{\rm v}$ [deg] & $50.20^{+3.85}_{-3.74}$ & $18.16^{+1.86}_{-1.48}$ \\
        $\log_{10}\epsilon_{\rm e}$ & $-1.49^{+0.42}_{-0.36}$ & $-4.13^{+0.95}_{-0.93}$ \\
        $\log_{10}\epsilon_{\rm B}$ & $-3.27^{+0.41}_{-0.35}$ & $-3.86^{+1.00}_{-0.91}$ \\
        $p$ & $2.12^{+0.01}_{-0.01}$ & $2.12^{+0.01}_{-0.01}$ \\
        $RA_0$ [MAS] & - & $-1.95^{+0.30}_{-0.29}$ \\
        $Dec_0$ [MAS] & - & $0.24^{+0.35}_{-0.33}$ \\
        $\alpha$ [rad] & - & $-0.02^{+0.11}_{-0.11}$ \\
	\hline
    \end{tabular}
\end{table}

\subsection{fitting results and discussions}

Our fitting results largely align with previous studies which also account for apparent superluminal motion (e.g., \citealt{2022Natur.610..273M,2024MNRAS.528.2600G,2023arXiv231002328R}). We observe notable differences in the parameter estimations between the two MCMC studies, particularly regarding the jet half-opening angle ($\theta_{\rm c}$) and the observing angle ($\theta_{\rm v}$). Specifically, when the centroid motion is excluded, the light curves are well-fitted, but this leads to a considerably large jet half-opening angle (7.°55) and observing angle (50.°2). Conversely, including all datasets results in a good fit for the centroid motion data, albeit with slight discrepancies in the light curves at later times. This approach yields smaller estimates for both the jet half-opening angle (2.°84) and the observing angle (18.°16). This finding is consistent with the aforementioned works.

The discrepancy observed between the two sets of fittings arises from the respective characteristics of the data sets used to deduce the jet half-opening angle and observing angle. In both approaches, the ratio $\theta_{\rm v}/\theta_{\rm c}$ is derived from the rising slope of the light curves, but the data features for determining the respective values differ between the two fittings. In the fitting that excludes the centroid motion, the jet half-opening angle is primarily inferred from the decay phase after the peak. During this phase, the jet transitions to a sub-relativistic speed, and the flux density reflects the jet's total energy. Conversely, before the peak, the flux density is determined by the isotropic equivalent energy of a localized jet region due to beaming effects. Therefore, the jet half-opening angle is deduced from the scale ratio between these two phases. In the fitting that incorporates centroid motion data, the observing angle is predominantly derived from the apparent transverse velocity, and the likelihood function associated with this measurement has a greater influence than that of the late-time light-curve data. Our fitting indicates that the two ways of inferring the jet half-opening angle and observing angles yield different estimation values.

This deviation may suggest a discrepancy between the late-time light curve and the observed superluminal motion. The late-time light curve's shallow decay suggests a wide jet, whereas the superluminal motion implies a narrow jet. However, using observation angles to infer the Hubble constant through joint fitting with gravitational wave data, previous studies (e.g., \citealt{2024MNRAS.528.2600G}) have shown that a narrow jet scenario aligns more closely with the current constraints set by Planck \citep{2020A&A...641A...6P} and SH0ES \citep{2022ApJ...934L...7R}. This result implies that a narrow jet scenario is more plausible than a wide jet. Thus, the discrepancy might indicate an excess at late times, potentially caused by an additional component, such as a kilonova afterglow.

While the late-time deviation is also noted in the aforementioned works, the excess observed in our results is slightly more significant. This can be attributed to our model, which predicts a steeper decline in narrow jets during the spreading phase in contrast to semianalytic methods (see \S \ref{subsec:afterglowpy} for a detailed discussion). This finding suggests that the jet in this event may be so narrow that its spreading phase is best represented by a hydrodynamic model.

This discrepancy might also arise from other factors which do not require an additional component. For instance, \citealt{2023arXiv231002328R} have utilized a Poisson likelihood function for the X-ray data, which significantly enhances the fitting results compared to the chi-square likelihood function. However, this approach is unable to account for the excess observed, as well in the radio wave band, and the radio dataset employed in this study is insufficient to span the late-time light curve. Additionally, certain nonstandard processes in GRB afterglows may also explain the shallow decay observed at late times, such as complex jet structures or changes in microphysical parameters during the evolution. A more thorough investigation that incorporates these systematics will be conducted in future research.

\section{Summary and Conclusion} \label{sec:summary}
In this work, we have developed a reduced hydrodynamic model for relativistic structured jets. The blast wave is approximated as an infinitely thin two-dimensional surface. By additionally assuming the axial symmetry of the jet, the simulation is effectively simplified into a one-dimensional representation. The radial integrate values are calibrated with the exact Blandford-McKee and Sedov-Taylor self-similar solutions in the respective limits. This is motivated by the effectively thin blast shell and the homogeneous shell approximation. Our method maintains the accuracy of numerical hydrodynamics, while attaining a computational efficiency that is comparable to semianalytic methods.

We have extensively compared the numerical results of our methods to existing tools. The comparison reveals a high level of agreement, especially with those from numerical hydrodynamics. We have also applied it to the case of GRB 170817A, fitting the light curve and flux centroid motion data. The parameter estimations mostly align with previous results, but our findings indicate a more evident late-time excess, suggesting the presence of an extra component or nonstandard processes. The possible existence of the additional component, which is also suggested in previous works, motivates a further investigation of GRB 170817A.

The methods discussed in this work have been developed into a numerical tool, which is called \texttt{jetsimpy}. The code is able to simulate a jet with arbitrary tabulated angular energy and Lorentz factor profiles, and can evolve the jet within an ISM environment, wind-like environment, or mixed environment. The simulation data is then used to calculate the light curve or spectrum, provided with a customized emissivity model. Additionally, the tool provides the calculation of the apparent superluminal motion and the image size of the jet. \texttt{jetsimpy} is optimized to be used in MCMC where millions of light curves with varying jet structures need to be generated in a reasonable time. It is also worth noting that although our model is built with GRB afterglow in mind, it is not limited to this scenario but to any relativistic hydrodynamic jets. As such, it will serve as a powerful tool for the community in the era of multimessenger astronomy.

\section*{Acknowledgments}
We thank Paz Beniamini for the useful discussion and suggestions. We also thank Ehud Narkar and Taya Govreen-Segal for providing simulation data that enables additional comparisons. Finally, we thank the anonymous referee for their careful review and valuable advice. H. W. and D. G. acknowledge support from the NSF AST-2107802, AST-2107806, and AST-2308090 grants.

\vspace{5mm}

\software{\texttt{astropy} \citep{astropy:2013,astropy:2018,astropy:2022}, \texttt{Matplotlib} \citep{Hunter:2007}, \texttt{Bilby} \citep{2019ApJS..241...27A}, \texttt{pymultinest}, \citep{2014A&A...564A.125B}}

\appendix

\section{The Numerical scheme} \label{appendix:scheme}

Here, we provide a concise overview of our numerical approach to solving hydrodynamic equations. The governing equations closely resemble those of regular numerical hydrodynamics, except that they are coupled with a Hamilton-Jacobi equation. 

\subsection{Eigenvalues of the Jacobian Matrix}
Similar to numerical hydrodynamics, our approach requires the eigenvalues of the Jacobian matrix to formulate the numerical scheme. In our model, due to the direct involvement of approximate trans-relativistic shock jump conditions, the eigenvalues are nontrivial and do not align with those found in regular relativistic hydrodynamics.

To begin with, we first address the solution of the Jacobian matrix. This is achieved by representing each component as a function of the conserved variables:
\begin{align}
    &\left|\frac{\partial {\bf F}}{\partial {\bf U}}\right| = \frac{c}{R}
    \begin{pmatrix}
    0 & 1 & 0 & 0\\
    (1-\beta_{\theta}^2)(1 + \frac{\partial P_{\rm sw}}{\partial E_{\rm b}}) - 1 & 2\beta_\theta & (1-\beta_{\theta}^2)\frac{\partial P_{\rm sw}}{\partial M_{\rm sw}} & (1-\beta_{\theta}^2)\frac{\partial P_{\rm sw}}{\partial M_{\rm ej}}\\
    -\beta_{\theta}\frac{M_{\rm sw}}{H_{\rm b}}(1 + \frac{\partial P_{\rm sw}}{\partial E_{\rm b}}) & \frac{M_{\rm sw}}{H_{\rm b}} & \beta_{\theta} - \beta_{\theta} \frac{M_{\rm sw}}{H_{\rm b}}\frac{\partial P_{\rm sw}}{\partial M_{\rm sw}} & -\beta_{\theta} \frac{M_{\rm sw}}{H_{\rm b}}\frac{\partial P_{\rm sw}}{\partial M_{\rm ej}}\\
    -\beta_{\theta} \frac{M_{\rm ej}}{H_{\rm b}}(1 + \frac{\partial P_{\rm sw}}{\partial E_{\rm b}}) & \frac{M_{\rm ej}}{H_{\rm b}} & -\beta_{\theta} \frac{M_{\rm ej}}{H_{\rm b}}\frac{\partial P_{\rm sw}}{\partial M_{\rm sw}} & \beta_{\theta} - \beta_{\theta} \frac{M_{\rm ej}}{H_{\rm b}}\frac{\partial P_{\rm sw}}{\partial M_{\rm ej}}
    \end{pmatrix}.
\end{align}
The four eigenvalues of this matrix can be analytically solved:
\begin{align}
    \omega_{1,2} & = \beta_{\theta}c/R, \\
    \omega_{3,4} & = \frac{\beta_{\theta}c}{R}\left(1 - \frac{1}{2}\frac{M_{\rm ej}}{H_{\rm b}}\frac{\partial P_{\rm sw}}{\partial M_{\rm ej}} - \frac{1}{2}\frac{M_{\rm sw}}{H_{\rm b}}\frac{\partial P_{\rm sw}}{\partial M_{\rm sw}}\right)\nonumber\\
    & \pm \frac{c}{R} \sqrt{(1-\beta_{\theta}^2)\left(\frac{\partial P_{\rm sw}}{\partial E_{\rm b}}+\frac{M_{\rm ej}}{H_{\rm b}}\frac{\partial P_{\rm sw}}{\partial M_{\rm ej}}+\frac{M_{\rm sw}}{H_{\rm b}}\frac{\partial P_{\rm sw}}{\partial M_{\rm sw}}\right)+\frac{1}{4}\beta_{\theta}^2\left(\frac{M_{\rm ej}}{H_{\rm b}}\frac{\partial P_{\rm sw}}{\partial M_{\rm ej}}+\frac{M_{\rm sw}}{H_{\rm b}}\frac{\partial P_{\rm sw}}{\partial M_{\rm sw}}\right)}.
\end{align}
The partial derivatives can be solved by differentiating eq. \ref{eq:energy_relation} and \ref{eq:pressure_relation}, yielding the following results:
\begin{align}
    \frac{\partial P_{\rm sw}}{\partial E_{\rm b}} & = \frac{\frac{2}{3}sM_{\rm sw}}{\frac{2}{3}sM_{\rm sw}(4\gamma^4-1)+(1-s)\gamma^3M_{\rm sw}+\gamma^3M_{\rm ej}}, \\
    \frac{\partial P_{\rm sw}}{\partial M_{\rm sw}} & = \frac{1}{3}s\beta^2 - \frac{\frac{2}{3}sM_{\rm sw}\left[s\gamma^2(1+\frac{1}{3}\beta^4)+(1-s)\gamma\right]}{\frac{2}{3}sM_{\rm sw}(4\gamma^4-1)+(1-s)\gamma^3M_{\rm sw}+\gamma^3M_{\rm ej}}, \\
    \frac{\partial P_{\rm sw}}{\partial M_{\rm ej}} & = - \frac{\frac{2}{3}s\gamma M_{\rm sw}}{\frac{2}{3}sM_{\rm sw}(4\gamma^4-1)+(1-s)\gamma^3M_{\rm sw}+\gamma^3M_{\rm ej}}.
\end{align}

The eigenvalues correspond to all the wave speeds present in this system. These values approach zero as $\beta_{\theta}\to 0$ and $\gamma\to\infty$. This effect suggests that, in the ultrarelativistic limit, the fluid elements are initially causally disconnected, with interactions across the blast surface initially frozen. In the numerical scheme we introduce in the following section, we do not require all the eigenvalues but only the maximum absolute value $\omega_{\rm max}=\max_{\rm j}(|\omega_j|)$. 

\subsection{Solving the evolution equations}
The evolution equations in our model are hyperbolic conservation laws coupled with a Hamilton-Jacobi equation. Each of them is solved with their respective numerical schemes.

To solve the hyperbolic equations, we employ a second-order Godunov-type finite volume method. The mesh is formed by unequally spaced (but static) cells arranged along the polar direction, as detailed in the following section. Up to second-order accuracy, the averages within each cell align with the values at the cell centers. This alignment is especially beneficial for our numerical scheme because the blast wave radius at each cell is also evaluated at the cell centers. The numerical scheme can be expressed in a semi-discrete form:
\begin{equation}\label{eq:semi-discrete}
    \frac{d {\bf U}_{\rm i}}{d t} = \frac{{\bf F}_{\rm i - 1/2}\sin\theta_{\rm i-1/2} - {\bf F}_{\rm i+1/2}\sin\theta_{\rm i+1/2}}{\Delta V_{\rm i}} + {{\bf S}}_{\rm i}.
\end{equation}
where $i$ denotes the center of a cell, and $i-1/2$ and $i+1/2$ are the left-hand and right-hand edges of the cell. The volume of each cell is $V_{\rm i}=\cos\theta_{\rm i-1/2}-\cos\theta_{\rm i+1/2}$. 

For the reconstruction of the primary variables within a cell, we perform a piecewise linear reconstruction approach with a Minmod slope limiter. The slopes for the cells adjacent to the poles are reconstructed using ghost cells, which are extrapolated based on a reflective boundary condition. The numerical flux at the cell edges is determined using the Rusanov Riemann solver  \citep{RUSANOV1962304}:
\begin{equation}
    {\bf F}_{\rm i+1/2} = \frac{1}{2}\left[{\bf F}_{\rm L} + {\bf F}_{\rm R} - \omega^{\rm stencil}_{\rm max}({\bf U}_{\rm R} - {\bf U}_{\rm L})\right]
\end{equation}
where ${\rm L}$ and ${\rm R}$ denote the values at the left-hand and right-hand sides of the edges. The coefficient $\omega^{\rm stencil}_{\rm max}$ is the maximum $\omega_{\rm max}$ over the four cells around the edge. The time integration is performed by a second-order strong stability preserving Runge-Kutta method with a time step $\Delta t$. 

To maintain numerical stability, the time steps must satisfy the following Courant–Friedrichs–Lewy (CFL) condition
\begin{equation}
    \Delta t = c_{\rm cfl}\max_{\rm i}\left(\frac{\theta_{\rm i+1/2}-\theta_{\rm i-1/2}}{\omega^{\rm sig}_{\rm i}}\right)
\end{equation}
where $0<c_{\rm cfl}<1$ is the CFL number and $\omega^{\rm sig}_{\rm i}$ is the signal speed of each cell. 

In principle the signal speed $\omega^{\rm sig}_{\rm i}$ should match $\omega_{\rm max}$. However, due to the aforementioned causality considerations, we find $\omega_{\rm max}\approx 0$ at very early times. This leads to a practically infinite time step at the beginning, which blows up the scheme. To ensure a finite time step, we apply the following correction:
\begin{equation}
    \omega^{\rm sig}_{\rm i} = \omega_{\rm max,i} + 0.05\frac{\beta_{\rm i}c}{R_{\rm i}}.
\end{equation}
In the ultrarelativistic limit, the correction term corresponds to a 5\% speed of light, and it vanishes in the Newtonian limit. In numerical tests, we find this correction provides an appropriate step size such that in the coasting phase the step size is sufficiently large to maintain good computational efficiency, while not so large that it misses the start of lateral expansion.

The radius evolution equation can be regarded as a nonlinear Hamilton-Jacobi equation $\partial R/\partial t+H(\partial R/\partial\theta)=0$, where the Hamiltonian $H$ is defined by the negative right-hand side of eq. \ref{eq:radial}. Following \citealt{shu2007high}, we adopt a second-order monotone scheme within the framework of the finite difference method
\begin{equation}
    \frac{d R}{d t} = \hat{H}(u^-, u^+),
\end{equation}
where $\hat{H}$ is the numerical Hamiltonian, and $u^-$ and $u^+$ are the left-hand and right-hand biased estimations of slope. The numerical Hamiltonian is calculated using a Lax-Friedrichs flux \citep{1989JCoPh..83...32S}:
\begin{equation}
    \hat{H}(u^-, u^+)=H(\frac{u^- + u^+}{2})-\frac{1}{2}\alpha_{\rm max}(u^+ - u^-),
\end{equation}
where $\alpha=|\partial H/\partial u|=|\beta_{\theta}c/R|$. The maximum is evaluated over a stencil including the cell itself and the two adjacent cells. The time integration is the same as the hyperbolic equations.

\subsection{Initial condition and resolution}

The initial conditions are defined by an arbitrary initial energy profile $E_{\rm b,0}$ and a Lorentz factor profile $\gamma_0$. Correspondingly, the ejecta mass profile is determined as $M_{\rm ej,0}=E_{\rm b, 0}/\gamma_0$. In a realistic scenario, a collimated jet has negligible energy far from the core. However, for numerical reasons, it is necessary to avoid 0 energy. This is achieved by adding an isotropic tail with sufficiently low energy and speed. To avoid zero radius as well, we initially ``coast" the jet from the origin to a sufficiently small radius at a constant initial speed, and start the simulation from then. The simulated data is subsequently interpolated for the calculation of the afterglow.

The resolution of the simulation depends on the initial condition. Cells are unequally sampled to adapt the energy and Lorentz factor distribution. For simplicity, we uniformly distribute the cells within the jet core, and logarithmic outside the core.

\section{Posterior Probability Distribution of the fitting to GRB 170817A}
\label{appendix:posterior}

\begin{figure*}[ht!]
\centering
    \includegraphics[width=\textwidth]{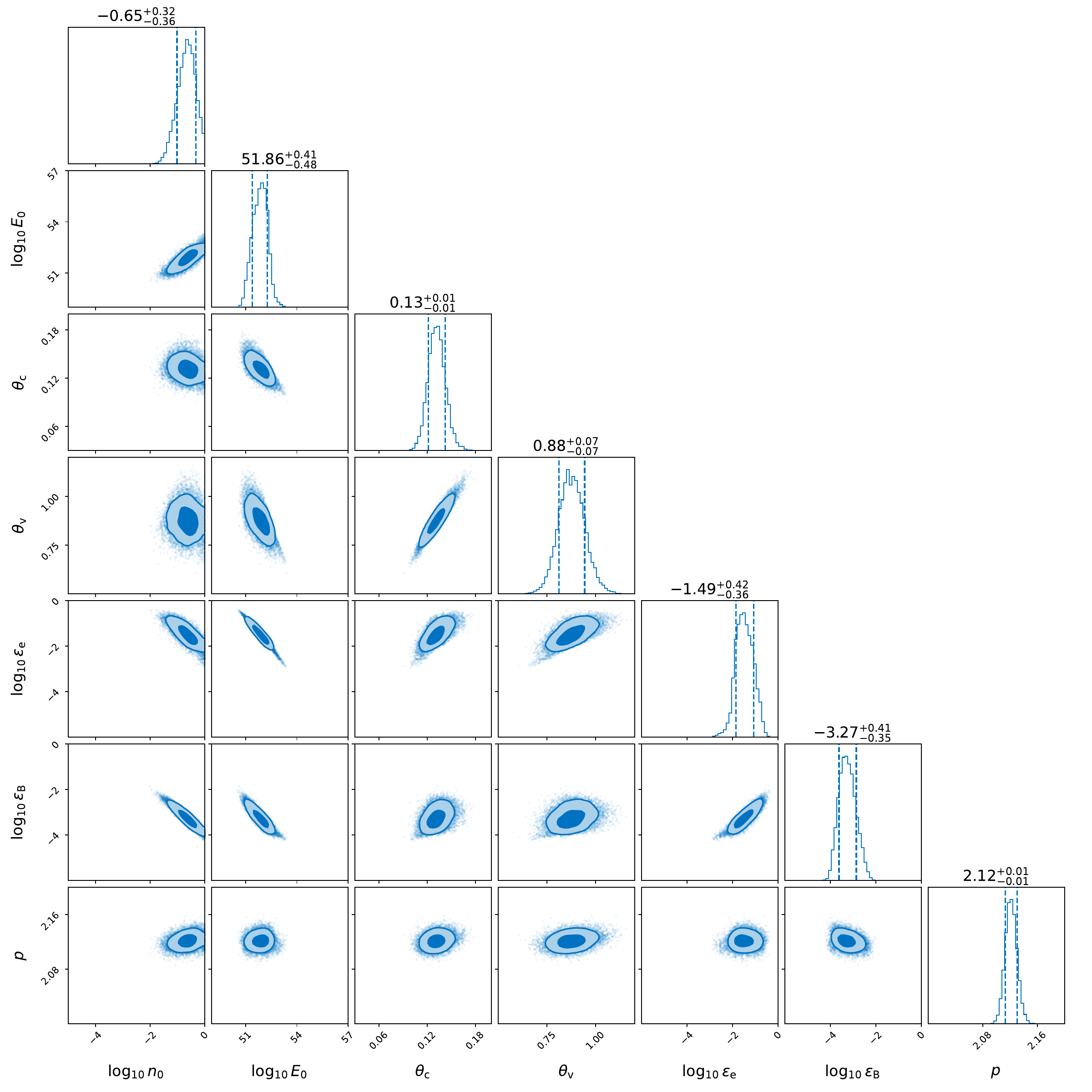}
\caption{The posterior probability distribution of the parameters in the MCMC study {\it without} the centroid offset data.
\label{fig:corner}}
\end{figure*}

\begin{figure*}[ht!]
\centering
    \includegraphics[width=\textwidth]{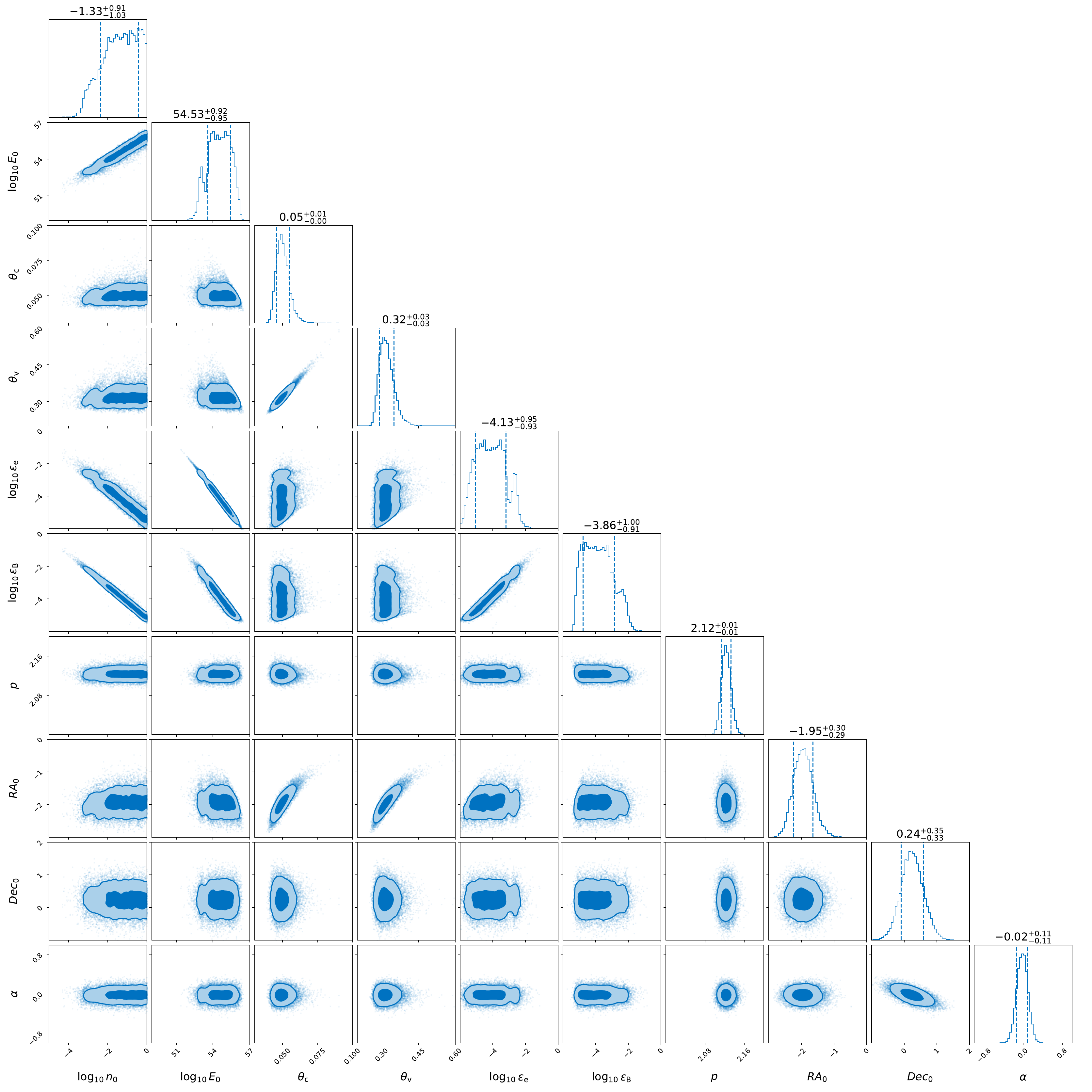}
\caption{The posterior probability distribution of the parameters in the MCMC study {\it with} the centroid offset data included.
\label{fig:corner_offset}}
\end{figure*}

\bibliography{sample631}{}
\bibliographystyle{aasjournal}

\end{CJK*}
\end{document}